\documentclass[prd,showpacs,superscriptaddress,twocolumn,
floatfix,preprintnumbers,altaffilletter]{revtex4}
\usepackage{graphicx,url,amssymb,amsmath,longtable,rotating,color,units,wasysym,subfigure,wrapfig,epsfig,multirow}

\begin{document}

\title{Correlated noise in networks of gravitational-wave detectors: \\
subtraction and mitigation}

\author{E. Thrane}
\address{LIGO Laboratory, California Institute of Technology, Pasadena, California 91125, USA}
\email{ethrane@ligo.caltech.edu}

\author{N. Christensen}
\address{Physics and Astronomy, Carleton College, Northfield, MN 55057, USA}
\email{nchriste@carleton.edu}

\author{R. M. S. Schofield}
\address{University of Oregon, Eugene, Oregon 97403, USA}
\email{rmssrmss@gmail.com}

\author{A. Effler}
\address{Louisiana State University, Baton Rouge, Louisiana 70803, USA}
\email{aeffle2@lsu.edu}

\begin{abstract}
  One of the key science goals of advanced gravitational-wave detectors is to observe a stochastic gravitational-wave background.
  However, recent work demonstrates that correlated magnetic fields from Schumann resonances can produce correlated strain noise over global distances, potentially limiting the sensitivity of stochastic background searches with advanced detectors.
  In this paper, we estimate the correlated noise budget for the worldwide advanced detector network and conclude that correlated noise may affect upcoming measurements.
  We investigate the possibility of a Wiener filtering scheme to subtract correlated noise from Advanced LIGO searches, and estimate the required specifications.
  We also consider the possibility that residual correlated noise remains following subtraction, and we devise an optimal strategy for measuring astronomical parameters in the presence of correlated noise.
  Using this new formalism, we estimate the loss of sensitivity for a  broadband, isotropic stochastic background search using $\unit[1]{yr}$ of LIGO data at design sensitivity.
% June 16:
%  Given our current noise budget, the uncertainty with which LIGO can estimate energy density will likely increase by a factor of $\approx$$4$---if it is impossible to achieve significant subtraction.
  Given our current noise budget, the uncertainty with which LIGO can estimate energy density will likely increase by a factor of $\approx$$12$---if it is impossible to achieve significant subtraction.
% June 16:
%  Additionally, narrowband cross-correlation searches may be severely affected at low frequencies $f\lesssim\unit[45]{Hz}$ without effective subtraction.
  Additionally, narrowband cross-correlation searches may be severely affected at low frequencies $f\lesssim\unit[70]{Hz}$ without effective subtraction.
\end{abstract}

\maketitle

\section{Introduction}
One of the most compelling targets for upcoming second-generation gravitational-wave detectors is the stochastic background.
Stochastic backgrounds can be created from the superposition of astrophysical sources such as binary coalescences~\cite{StochCBC}, magnetars~\cite{cutler,WuMagnetars}, rotating neutron stars~\cite{RegPac,owen,barmodes1,barmodes2,barmodes3}, and the first stars~\cite{firststars}.
Cosmological signals may arise during or following inflation~\cite{eastherlim,peloso}, from cosmic strings~\cite{caldwellallen,DV1,DV2,cosmstrpaper,olmez1,olmez2}, phase transitions~\cite{durrer}, and from non-standard cosmologies~\cite{PBB1,PBBpaper}.
Recent observations by BICEP2 of a gravitational-wave background at very low frequencies, if confirmed, represent the first observational signature of a cosmological background and a remarkable test of inflation~\cite{bicep2}.

Over the coming years, a worldwide network of gravitational-wave detectors~\cite{aLIGO2,aVirgo,GEO2,kagra} will probe gravitational-wave energy density several orders of magnitude below the current best limits: $\Omega_\text{gw}<6.9\times10^{-6}$ at 95\% confidence in a band around $\unit[100]{Hz}$~\cite{stoch-S5}.
By measuring the stochastic background over a wide range of frequencies, we may gain insights into important epochs in the history of the universe~\cite{maggiore}.

Searches for the stochastic background rely on the principle of cross-correlation~\cite{allen-romano,christensen_prd}.
By integrating over a data period of about a year, it is possible to dig far below the detectors' nominal strain noise.
A key premise in past cross-correlation searches is that the noise in each detector is uncorrelated.
Correlated noise creates a systematic bias, which is not reduced with continued integration.
In a previous paper, we showed that magnetic fields from global Schumann resonances~\cite{schumann,schumann-b} can create correlated noise in a global network of gravitational-wave detectors~\cite{schumann_ligo}.
While correlated noise from Schumann resonances is too low-level to affect stochastic searches with first-generation detectors, we warned that it might be significant for advanced detectors~\cite{schumann_ligo}.
(See~\cite{Fotopoulos} for a discussion of correlated noise in the context of colocated detectors.)

In this paper we provide an updated correlated noise budget for the Advanced LIGO detector network, consisting of detectors in Hanford, WA and Livingston, LA.
The advanced detector network will also include Virgo, KAGRA, and possibly LIGO India.
We focus here on the two LIGO detectors, which are expected to provide the most sensitive stochastic background measurement in the near future.
However, we note that correlated noise from Schumann resonances is a concern for all detector pairs in this worldwide network.

We show that correlated noise may be significant, and so we investigate the possibility of subtracting the correlated noise with a Wiener filtering scheme.
We use a toy model to demonstrate some of the general features and the limitations of Wiener filtering.
Then, we perform a numerical study to evaluate quantitatively the prospects for effective subtraction in advanced detector networks.
Schumann fields lurk at amplitudes below typical anthropogenic magnetic noise levels, making subtraction a non-trivial proposition.
We thus investigate how best to handle residual correlated noise if it cannot be entirely subtracted.

The remainder of this paper is organized as follows.
In Section~\ref{definitions}, we introduce formalism that will be useful for our discussion of correlated noise.
In Section~\ref{measure}, we  present a {\em preliminary} measurement-based correlated noise budget for the Advanced LIGO network.
This preliminary noise budget is sure to be updated in the coming months and years as the detector is continually commissioned.
We expect our preliminary noise budget to provide an approximately accurate forecast for the foreseeable future up to a factor of a few.
In Section~\ref{wiener}, we describe how Wiener filtering can be used to reduce correlated noise.
Through analytical calculations and numerical studies, we determine the key ingredients for successful subtraction while highlighting potential pitfalls.

In Section~\ref{aligo}, we apply the lessons learned in Section~\ref{wiener} to the problem of subtracting correlated magnetic noise from Schumann resonances in a network of gravitational-wave detectors.
We carry out a systematic numerical study in order to estimate the effectiveness of subtraction given different levels of magnetic coupling.
In Section~\ref{contingency}, we consider how best to proceed if correlated noise remains following subtraction.
We elucidate the optimal strategy for measuring astrophysical parameters in the presence of residual correlated noise, and we show how cross-correlation searches are affected.
We offer concluding remarks in Section~\ref{conclusions}.

\section{Definitions and formalism}\label{definitions}
We consider two strain channels $s$, which contain an astrophysical component $h$, uncorrelated noise $n$, and correlated noise $m$, which couples to each detector through a transfer function $r$:
\begin{equation}\label{eq:s1s2}
  \begin{split}
    \tilde{s}_1(f) = \tilde{h}_1(f) + \tilde{n}_1(f) + r_1(f) \, \tilde{m}(f) \\
    \tilde{s}_2(f) = \tilde{h}_2(f) + \tilde{n}_2(f) + r_2(f) \, \tilde{m}(f)
  \end{split}
\end{equation}
Here, $f$ is frequency and tildes denote Fourier transforms.
We make the following assumptions:
\begin{equation}\label{eq:defs}
  \begin{split}
    k\, \langle\tilde{h}^*_1(f) \tilde{h}_2(f)\rangle \equiv H(f) \neq 0 \\
    %------------------------------------------------------------------
    \langle\tilde{h}^*_1(f) \tilde{n}_2(f)\rangle =
    \langle\tilde{n}^*_1(f) \tilde{h}_2(f)\rangle = 0 \\
    %------------------------------------------------------------------
    \langle\tilde{m}^*(f) \tilde{h}_2(f)\rangle =
    \langle\tilde{h}^*_1(f) \tilde{m}(f)\rangle = 0 \\
    %------------------------------------------------------------------
    \langle\tilde{n}^*_1(f) \tilde{m}(f)\rangle=
    \langle\tilde{m}^*(f) \tilde{n}_2(f)\rangle = 0 \\
    %------------------------------------------------------------------
    \langle\tilde{n}^*_1(f) \tilde{n}_2(f)\rangle = 0 \\
    k\, \langle\tilde{m}^*(f) \tilde{m}(f)\rangle \equiv M(f) \neq 0 .
  \end{split}
\end{equation}
$H(f)$ is the {\it astrophysical} strain cross-power spectrum that we seek to measure, $M(f)$ is the correlated {\it magnetic} noise, and $k$ is a Fourier normalization constant.
Angled brackets denote the expectation value averaged over many trials.

When the correlated noise is negligible,  $H(f)$ can be estimated as a time averaged cross-power
\begin{equation}\label{eq:Y}
  \begin{split}
    \widehat{Y}(f) & \equiv k\,
    \text{Re} \left[ \overline{\tilde{s}^*_1(f) \tilde{s}_2(f)} \right] \\
    & \equiv k\,
    \text{Re} \left[\frac{1}{N_\text{segs}}\sum_{t=1}^{N_\text{segs}} \tilde{s}^*_1(t;f) \tilde{s}_2(t;f) \right] ,
  \end{split}
\end{equation}
with associated uncertainty estimated by~\cite{stoch-S5,locus}:
\begin{equation}\label{eq:sigma}
  \sigma^2(f) = \frac{1}{2} \frac{1}{N_\text{segs}} \overline{P_1(f) P_2(f)} .
\end{equation}
Here, $(t;f)$ indicates the discrete Fourier transform of a (typically $\unit[60]{s}$-long) data segment beginning at time $t$ and $N_\text{segs}$ is the total number of segments in the observing period (typically $\unit[1]{yr}$).
The variables $P_1(f)$ and $P_2(f)$ are the noise (auto) power spectra.
We assume that, in each segment, the uncorrelated noise dominates over both the astrophysical signal and the correlated noise so that
\begin{equation}
  \begin{split}
    P_1(f) & \equiv k\, \langle\left| \tilde{s}_1(f) \right|^2 \rangle \approx
    k\, \langle \left| \tilde{n}_1(f) \right|^2 \rangle \\
    P_2(f) & \equiv k\, \langle \left| \tilde{s}_2(f) \right|^2 \rangle \approx
    k\, \langle \left| \tilde{n}_2(f) \right|^2 \rangle .
  \end{split}
\end{equation}
For the sake of simplicity, we assume that the noise in each detector is comparable so that $P_1(f)=P_2(f)\equiv P(f)$.

The {\em correlated strain noise} power spectrum is:
\begin{equation}\label{eq:HM}
  H_M(f) \equiv \text{Re} \left[ r^*_1(f) r_2(f) M(f) \right] .
\end{equation}
While the statistical uncertainty associated with $\widehat{Y}(f)$ falls like $N_\text{segs}^{-1/2}$, $H_M(f)$ represents a systematic error, which does not decrease with continued integration.
If it is non-negligible, it may be either subtracted or treated as a source of systematic uncertainty.

We have already defined several power spectra: $H(f)$, $M(f$), $P(f)$, and $H_M(f)$.
In the course of our investigations, it will be useful to define several more.
To help keep track, we have provided a summary of different power spectra in Tab.~\ref{tab:power}.

While Eq.~\ref{eq:HM} describes the correlated strain noise in a pair of detectors, it does not take into account the response of the detector pair to a stochastic background.
It is therefore useful to define a pair of estimators that do:
\begin{equation}\label{eq:OmegaM}
  \begin{split}
    \widehat{\Omega}_M(f|\alpha) & = \frac{2\pi^2}{3 H_0^2} \, f^{3-\alpha} \, \frac{5}{\gamma(f)} \, \widehat{H}_M(f) \\
    \sigma(f|\alpha) & = \frac{2\pi^2}{3 H_0^2} \, f^{3-\alpha} \frac{5}{|\gamma(f)|} \, \sigma(f) .
  \end{split}
\end{equation}
Here, $\gamma(f)$ is the overlap reduction function for the LIGO Hanford-Livingston detector pair~\cite{christensen_prd,allen-romano}, $H_0$ is the Hubble constant, and $\alpha$ is a spectral index.
When $\alpha=0$, $\widehat\Omega_M$ is an estimator for the {\em apparent} gravitational-wave energy density spectrum, which is, in reality, due to correlated noise.

The observable we actually seek to measure is gravitational-wave energy density:
\begin{equation}\label{eq:Omega}
  \Omega_\text{gw}(f) \equiv 
  \frac{1}{\rho_c}\frac{d\rho_\text{gw}}{\ln f} .
\end{equation}
Here, $\rho_\text{gw}$ is the energy density of gravitational waves and $\rho_c$ is the critical energy density for a flat universe.

When $\alpha\neq0$, then $\widehat\Omega_M(f|\alpha)$ does not have a simple physical interpretation, but it is still useful for optimally combining measurements across the observing band given an assumed spectral shape:
\begin{equation}
  \Omega_\text{gw}(f) \propto f^\alpha .
\end{equation}
We therefore define broadband estimators:
\begin{equation}\label{eq:broadband}
  \begin{split}
    \widehat\Omega_M(\alpha) & = \frac{\int df \, \widehat\Omega_M(f|\alpha) \,
      \sigma^{-2}(f|\alpha)} {\int df \, \sigma^{-2}(f|\alpha)} \\
    \sigma(\alpha) & = \left[ \int df \, \sigma^{-2}(f|\alpha) \right]^{-1/2}
  \end{split}
\end{equation}
The average signal-to-noise ratio {\em due to contamination from correlated noise} is:
\begin{equation}\label{eq:SNRM}
  \text{SNR}_M(\alpha) = \langle \widehat\Omega_M(\alpha) \rangle / \sigma(\alpha) .
\end{equation}
When $\text{SNR}_M(\alpha)\ll1$, then correlated noise can be safely ignored.
When $\text{SNR}_M(\alpha)\gtrsim1$, it must be taken into account.

\section{Noise budget}\label{measure}
In this section, we construct a correlated noise budget for Advanced LIGO.
This represents a first pass based on measurements using components of the still-to-be commissioned Advanced LIGO detectors.
While subsequent measurements are necessary to produce a more accurate and more complete noise budget, we expect this first pass to provide a reasonable rough estimate.

First, we estimate the magnetic-field-to-strain transfer function at each detector $r_1$, $r_2$.
We assume that the correlated magnetic field couples into the strain channels with a parameterized coupling function (approximately the same for both detectors).
The displacement of a single test mass in the presence of a magnetic field is approximately:
\begin{equation}
  \left|r(f)\right| = \kappa\,(\unit[4\times10^{-8}]{m \, T^{-1}})
  \left(f/\unit[10]{Hz}\right)^{-\beta} .
\end{equation}
The normalization $\kappa$ and the spectral index $\beta$ are estimated using a coil to inject large magnetic fields in the vicinity of LIGO test masses~\cite{alog}.
This translates into a magnetic-field-to-strain coupling function given by
\begin{equation}\label{eq:kappa_beta}
  % https://alog.ligo-la.caltech.edu/aLOG/index.php?callRep=11217
  \left|r(f)\right| = \kappa\,(\unit[10^{-23}]{strain \, pT^{-1}})
  \left(f/\unit[10]{Hz}\right)^{-\beta} .
\end{equation}
%(While one might hope that long-wavelength magnetic fields would couple primarily to the interferometers' common modes, magnetic field lines are dramatically distorted by the ubiquitous presence of metal~\cite{schumann_ligo}, and the magnetic coupling to each test mass is sure to be different, which all but ensures significant differential-mode contamination.)

The coupling function for a coil injection actuating on a single test mass is not identical to the coupling function for a site-wide Schumann resonance actuating on every test mass simultaneously.
While one might hope that long-wavelength magnetic fields would couple primarily to the interferometers' common modes, we expect significant differential-mode contamination for a number of reasons.

First, there is a differential component due to the different orientations of the X and Y arms, which experience different forces.
%Second, Advanced LIGO employs pairs of opposing magnets in order to minimize coupling, and so the observed single-test-mass coupling is indicative of random mismatch expected to be different for each test mass.
%Third, aside from the permanent magnets, coupling to ferromagnetic parts is not expected to be the same at each test mass because it depends on manufacture and field history.
%Fourth, for cable coupling, the effect on a particular test mass depends on non-cancelation of loops in cables that are twisted to reduce magnetic coupling, clearly different at the different masses.
Second, techniques are employed to minimize magnetic coupling and so residual coupling depends on, for example, mismatches in the strengths of oppositely oriented magnets that are intended to cancel each other, and other random effects that can alter the direction of forces on different test masses in the same field.
Finally, large ferromagnetic parts and rebar in the floor are expected to distort the direction and gradients of the field differently at each of the four test masses.

%However, we expect the two to agree to within a factor of $\approx$$2$.
%The differential strain can be estimated approximately by adding in quadrature the coupling of the four test masses and dividing by two because gravitational-wave strain is measured using only differential motion.
Given these complications, the differential strain can be estimated approximately by adding in quadrature the coupling of the four test masses and dividing by two because gravitational-wave strain is measured using only differential motion.
We expect this approximation to be valid to within a factor of $\approx$$2$.

In practice, the simple power-law parameterization of Eq.~\ref{eq:kappa_beta} is a simplification.
We expect the coupling function to include components from multiple coupling
sites.
At each of these sites, the coupling has a frequency dependence determined by the coupling mechanism and the filter poles associated with shielding.
Thus, coupling to magnets in the test mass suspension is expected to increase
approximately as $1/f^2$ below the shielding frequency that is associated with the vacuum chamber (about $\unit[20]{Hz}$) and to decrease as $1/f^3$ above that.
Coupling to unshielded signal cables is expected to increase as $f$. 
Thus, the total coupling function may have a complicated shape.
However, at present, we find that the measured coupling functions can be approximated as power laws.

The two scenarios we consider are: coupling to the length degree of freedom and indirect coupling through angular motion.
% June 16:
%We observe coupling to the length degree of freedom characterized by $\kappa=1$, $\beta=2.67$.
We observe coupling to the length degree of freedom characterized by $\kappa=2$, $\beta=2.67$.
We also observe coupling to the angular degree of freedom.
Angular noise couples to length noise when the interferometer beam is not aligned with the axis of rotation for each optic.
Assuming a plausible beam offset of $\unit[1]{mm}$, we obtain $\kappa=0.25$, $\beta=1.74$.
However, a beam offset of $\unit[3]{mm}$ would not be surprising, leading to  $\kappa=0.75$, $\beta=1.74$.

Next, we combine our coupling function measurements to previous measurements of correlated magnetic fields at the LIGO Hanford and Livingston detectors~\cite{schumann_ligo} in order to calculate $\widehat\Omega_M(f|\alpha)$; see Eq.~\ref{eq:OmegaM}.
We conservatively assume that the transfer function phase is such that the contamination from magnetic fields is maximal.
We consider three spectral indices: $\alpha=0$ (expected for a cosmological source~\cite{stoch-S5}), $\alpha=2/3$ (expected for an astrophysical background of binary coalescences~\cite{StochCBC}), and $\alpha=-2$ (chosen conservatively to emphasize the low frequency range where contamination is expected to be worst).

\begin{figure}[hbtp!]
  \begin{tabular}{c}
    \psfig{file=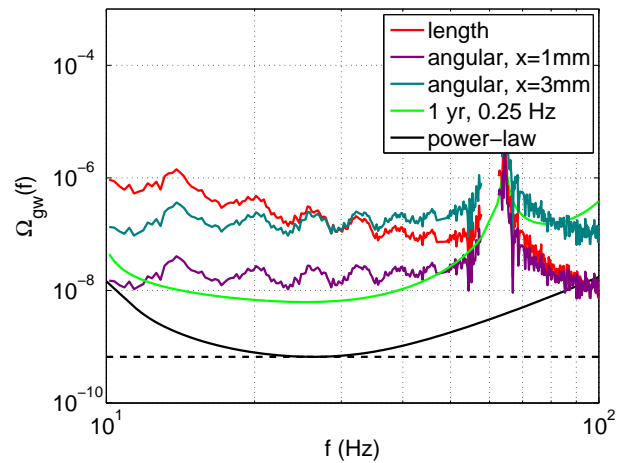, width=3.2in}
  \end{tabular}
  \caption{
    Correlated noise budget for a stochastic background search with the LIGO Hanford and Livingston detectors.
    The noise budget is given in terms of normalized energy density; see Eq.~\ref{eq:Omega}.
    Red is the correlated noise from coupling to the length degree of freedom.
    Purple and turquoise show the correlated noise from indirect coupling to the angular degree of freedom assuming a beam offset of $\unit[1]{mm}$ and $\unit[3]{mm}$ respectively.
    The green curve shows the statistical uncertainty obtained through $\unit[1]{yr}$ of integration at design sensitivity with a frequency resolution of $\unit[0.25]{Hz}$.
    (The peak just above $\unit[60]{Hz}$ is due to a zero in the overlap reduction function.)
    The black curve is the power-law integrated curve~\cite{locus}, which represents the broadband sensitivity of the search to an isotropic background with a power-law spectrum.
    The dashed black line is the sensitivity to a signal with flat energy density spectrum $\alpha=0$.
    Correlated noise may be safely ignored only if it falls below the power-law integrated curve.
    Note that the estimates shown here are based on current measurements, and the noise budget may change with continued commissioning.
    Coupling measurements are uncertain to a factor of $\approx$$2$, and we conservatively assume that the transfer function phase is such that contamination is maximal.
  }
  \label{fig:noisebudget}
\end{figure}

The noise budget for the Advanced LIGO Hanford-Livingston stochastic search is summarized in Fig.~\ref{fig:noisebudget}.
We include correlated noise from coupling to the length degree of freedom (red) as well as correlated noise coupling indirectly through angular motion.
%The latter depends on the offset in the beam from the optic center $x$, and so we include estimates for $x=\unit[1]{mm}$ (purple) and $x=\unit[3]{mm}$ (turquoise).
The latter depends on the offset of the beam from the optic's axis of rotation $x$, and so we include estimates for $x=\unit[1]{mm}$ (purple) and $x=\unit[3]{mm}$ (turquoise).
These correlated noise curves can be compared to the statistical uncertainty obtained from $\unit[1]{yr}$ of integration and using $\unit[0.25]{Hz}$ wide frequency bins (green).
By integrating over all frequency bins in the green curve, we obtain the black power-law integrated curve~\cite{locus}, which represents the sensitivity of the LIGO network to any isotropic stochastic background described by a power law.
(The dashed black line shows the sensitivity to a stochastic signal with a flat energy density spectrum $\alpha=0$.)
The correlated noise can be ignored only if it falls below the black power-law integrated curve.
The expected correlated noise is well above the power-law integrated curve, and so correlated noise cannot at present be ignored.

In order to quantify the extent of expected contamination, we calculate $\text{SNR}_M(\alpha)$ for the length and angular coupling scenarios.
The results are given in Tab.~\ref{tab:contamination}.
Correlated noise, coupling through the angular degree of freedom, is currently projected to induce a signal-to-noise ratio of $\text{SNR}_M(\alpha)\approx24$--$30$ depending on the spectral index $\alpha$ and assuming a beam offset of $x=\unit[1]{mm}$.
$\text{SNR}_M(\alpha)$ scales like $x^2$, so these numbers become $220$--$270$ for $x=\unit[3]{mm}$.
% June 16:
%Correlated noise, coupling to the length degree of freedom, is expected to induce a signal-to-noise ratio $\text{SNR}_M(\alpha)\lesssim82$--$120$.
Correlated noise, coupling to the length degree of freedom, is expected to induce a signal-to-noise ratio $\text{SNR}_M(\alpha)\lesssim330$--$470$.

\begin{table*}
  \begin{tabular}{|c|c|c|c|}
    \hline
% first version:
%    $2/3$ & 17 \\\hline
%    $0$ & 25 \\\hline
%    $-2$ & 48 \\\hline
% second version
%    $2/3$ & 7 \\\hline
%    $0$ & 9 \\\hline
%    $-2$ & 13 \\\hlin
    {\bf coupling} & {\bf $(\kappa,\beta)$} & {\bf spectral index $\alpha$} & {\bf $|\text{SNR}_M(\alpha)|$} \\\hline
    \multirow{3}{*}{angular ($\unit[1]{mm}$ beam offset)}
    & \multirow{3}{*}{$(0.25, 1.74)$}
    % may 20: updated after sqrt(2) fix
%    & $2/3$ & 21  \\
%    & &  $0$ & 21 \\
%    & & $-2$ & 17 \\\hline
    & $2/3$ & 29  \\
    & &  $0$ & 30 \\
    & & $-2$ & 24 \\\hline
    \multirow{3}{*}{angular ($\unit[3]{mm}$ beam offset)}
    & \multirow{3}{*}{$(0.75, 1.74)$}
    % may 20: updated after sqrt(2) fix
%    & $2/3$ & 190  \\
%    & &  $0$ & 190 \\
%    & & $-2$ & 150 \\\hline
    & $2/3$ & 260  \\
    & &  $0$ & 270 \\
    & & $-2$ & 220 \\\hline
    \multirow{3}{*}{length}
% June 16:
%    & \multirow{3}{*}{$(1, 2.67)$}
    & \multirow{3}{*}{$(2, 2.67)$}
    % may 20: updated after sqrt(2) fix
%    & $2/3$ & 56 \\
%    & &  $0$ & 67 \\
%    & & $-2$ & 82 \\\hline
% June 16: updated after Robert's post-processing factor of two
%    & $2/3$ & 82 \\
%    & &  $0$ & 95 \\
%    & & $-2$ & 120 \\\hline
    & $2/3$ & 330 \\
    & &  $0$ & 380 \\
    & & $-2$ & 470 \\\hline
    % numbers generated with snr_check.m and noisebudget.m.
  \end{tabular}
  \caption{
    The expected broadband contamination from correlated noise for coupling to the length degree of freedom and the angular degree of freedom assuming a beam offsets of $x=\unit[1]{mm}$ and  $x=\unit[3]{mm}$.
    The second column describes the coupling function with parameters $(\kappa,\beta)$; see Eq.~\ref{eq:kappa_beta}.
    The third column lists different spectral shapes for stochastic gravitational-wave backgrounds.
    The forth column gives the projected signal-to-noise ratio induced from correlated noise, assuming $\unit[1]{yr}$ of coincident data from the LIGO Hanford and Livingston detectors operating at design sensitivity.
  }
  \label{tab:contamination}
\end{table*}

It follows from Tab.~\ref{tab:contamination} that correlated noise is a serious concern for stochastic background searches with advanced detectors.
In the following sections we investigate strategies for subtraction and mitigation of correlated noise.

\section{Subtraction with Wiener filtering}\label{wiener}
In this section, we discuss how to subtract the correlated strain noise term $H_M(f)$.
The goal is to provide a foundational understanding for the numerical calculations presented in Section~\ref{aligo}.
However, our presentation here is somewhat general so that our results might be useful to a broad readership.

\subsection{Wiener filtering}
Wiener filtering~\cite{wiener} is a means of subtracting noise from a channel $s$ using a witness channel $w$ when the transfer function from $w$ to $s$ is not known a priori.
See~\cite{ligo_wiener} for a broad discussion of Wiener filtering in gravitational-wave detectors and~\cite{meadors,driggers,dmass} for recent examples.
For our application, we consider two witness sensors $w_1$ and $w_2$, which are used to measure the correlated noise $m$ present in (LIGO strain) channels $s_1$ and $s_2$.
In reality, magnetic fields are vector quantities, and so at least three sensors should be employed at each detector in order to completely characterize the Schumann resonance noise~\footnote{In theory, the vertical component of Schumann resonance magnetic fields is zero, but local distortions change this.}.
For the sake of simplicity, however, we work with individual sensors here, and return to the topic of multiple witness sensors in Subsection~\ref{multiple}.

In the time domain,
\begin{equation}\label{eq:w12}
  \begin{split}
    w_1(t) & = m(t) + \eta_1(t) \\
    w_2(t) & = m(t) + \eta_2(t) .
  \end{split}
\end{equation}
The $\eta$ terms represent the noise in each witness sensor, which limits the ability to accurately measure $m$.
This limiting noise may be instrumental (e.g., electronic) or environmental (e.g., from local magnetic fields).
By definition, $\langle \tilde\eta^*_1(f) \tilde\eta_2(f) \rangle=0$.
We henceforth assume that the witness sensor auto-power noise ${\cal N}(f)$ is the same in both channels: ${\cal N}(f) \equiv k\, \left|\tilde\eta_1(f)\right|^2 = k\, \left|\tilde\eta_2(f)\right|^2$.

The $w$-to-$s$ transfer function can be estimated as
\begin{equation}\label{eq:r1r2}
  \begin{split}
    \widehat{r}_1(f) \equiv \frac{\overline{\tilde{s}_1(f) \tilde{w}_1^*(f)}}
    {\overline{\left|\tilde{w}_1(f)\right|^2}} \\
    \widehat{r}_2(f) \equiv \frac{\overline{\tilde{s}_2(f) \tilde{w}_2^*(f)}}
    {\overline{\left|\tilde{w}_2(f)\right|^2}} ,
  \end{split}
\end{equation}
where the overline denotes time-averaging.
The hats on $\widehat{r}_1$ and $\widehat{r}_2$ denote that they are estimated quantities; the true transfer functions are denoted $r_1$ and $r_2$ with no hats; see Eq.~\ref{eq:s1s2}.

The Wiener-subtracted data are given by
\begin{equation}\label{eq:wiener}
  \begin{split}
    \tilde{s}'_1(f) & = \tilde{s}_1(f) - \widehat{r}_1(f) \tilde{w}_1(f) \\
    \tilde{s}'_2(f) & = \tilde{s}_2(f) - \widehat{r}_2(f) \tilde{w}_2(f) .
  \end{split}
\end{equation}
We refer to $s_1$ and $s_2$ (and estimators calculated using them) as ``dirty'' whereas $s'_1$ and $s'_2$ (and associated estimators) are ``clean.''

It is constructive to consider the expectation value of the clean version of Eq.~\ref{eq:Y}:
\begin{widetext}
  \begin{equation}\label{eq:s1ps2p}
    \langle \widehat{Y}'(f) \rangle \equiv
    k\, \text{Re}\Big[ \langle \tilde{s}'^*_1(f) \tilde{s}'_2(f) \rangle\Big] = 
    H(f) + \text{Re}\Bigg[ 
      r_1^*(f) r_2(f) M(f) + \Big(
    \widehat{r}_1^*(f) \widehat{r}_2(f)
    -r_1^*(f) \widehat{r}_2(f)
    -\widehat{r}_1^*(f) r_2(f)
    \Big) M(f) \Bigg] .
  \end{equation}
\end{widetext}
In the limit that $\widehat{r_1}=r_1$ and $\widehat{r_2}=r_2$,
\begin{equation}
  \langle \widehat{Y}'(f) \rangle \rightarrow H(f) .
\end{equation}
In this limit, there is no correlated noise; the subtraction is perfect.

In reality, $\widehat{r_1}\neq r_1$ and $\widehat{r_2} \neq r_2$, and this is a key limitation of Wiener filtering in practice.
We gain further insight if we consider the ratio of the expectation values of the numerator and denominator of a generic Wiener filter $\widehat{r}$:
\begin{equation}
  \frac{\langle \tilde{s}(f) \tilde{w}^*(f) \rangle}
       {\langle \left|\tilde{w}(f)\right|^2 \rangle}
       = 
       \frac{r(f) M(f)}{M(f)+{\cal N}(f)} .
\end{equation}
In the limit that $M(f)\gg{\cal N}(f)$, this ratio becomes
\begin{equation}\label{eq:goes_to_r}
  \frac{\langle \tilde{s}(f) \tilde{w}^*(f) \rangle}
       {\langle \left|\tilde{w}(f)\right|^2 \rangle}
       \rightarrow r(f) ,
\end{equation}
and Wiener filtering is successful.
In the opposite limit that $M(f)\ll{\cal N}(f)$, the ratio becomes
\begin{equation}\label{eq:goes_to_zero}
  \frac{\langle \tilde{s}(f) \tilde{w}^*(f) \rangle}
       {\langle \left|\tilde{w}(f)\right|^2 \rangle}
       \rightarrow 0 ,
\end{equation}
and Wiener filtering fails.

From these considerations, it is useful to define ``witness signal-to-noise ratio'' (not to be confused with the broadband signal-to-noise ratio due to contamination from correlated noise defined in Eq.~\ref{eq:SNRM}):
\begin{equation}\label{eq:rho_m}
  \rho_w^2(f) \equiv M(f) / {\cal N}(f) .
\end{equation}
Any proposal for Wiener filtered subtraction must ensure that $\rho_w(f)$ is sufficiently high to provide suitable subtraction in the band of interest.

\subsection{Numerical illustration}\label{numerical}
Numerical simulations are useful for demonstrating the qualitative behavior of Wiener subtraction.
In this section we consider a toy model.
While it is simple, it serves to illustrate salient features.

We simulate data for four channels: $s_1$, $s_2$, $w_1$, and $w_2$.
The two signal channels ($s_1$ and $s_2$) consist of uncorrelated Gaussian white noise plus a small amount of correlated noise ($10\%$ of the power of the uncorrelated noise).
The correlated noise is ``measured'' by witness channels ($w_1$ and $w_2$) with a witness signal-to-noise ratio $\rho_w$ (see Eq.~\ref{eq:rho_m}).
The witness noise is also assumed to be white, and so $\rho_w$ is independent of frequency.

In order to evaluate the success of Wiener filtering, we calculate the coherence $\text{coh}(f)$ between strain channels $1$ and $2$
\begin{equation}\label{eq:coh}
  \begin{split}
    \text{coh}(f) \equiv
    \frac{\left|\overline{\tilde{s}_1^*(f) \tilde{s}_2(f)}\right|^2}
         {\overline{\left|\tilde{s}_1(f)\right|^2} \,
           \overline{\left|\tilde{s}_2(f)\right|^2} } \quad \text{(dirty)} \\
    \text{coh}(f) \equiv
    \frac{\left|\overline{\tilde{s}_1^{'*}(f) \tilde{s}'_2(f)}\right|^2}
         {\overline{\left|\tilde{s}'_1(f)\right|^2} \,
           \overline{\left|\tilde{s}'_2(f)\right|^2} } \quad \text{(clean)} \\
  \end{split}
\end{equation}
to study how the cleaned coherence varies as a function of $\rho_w$.
As before, the overline denotes time-averaging.

\begin{figure*}[hbtp!]
  \begin{tabular}{cc}
    \psfig{file=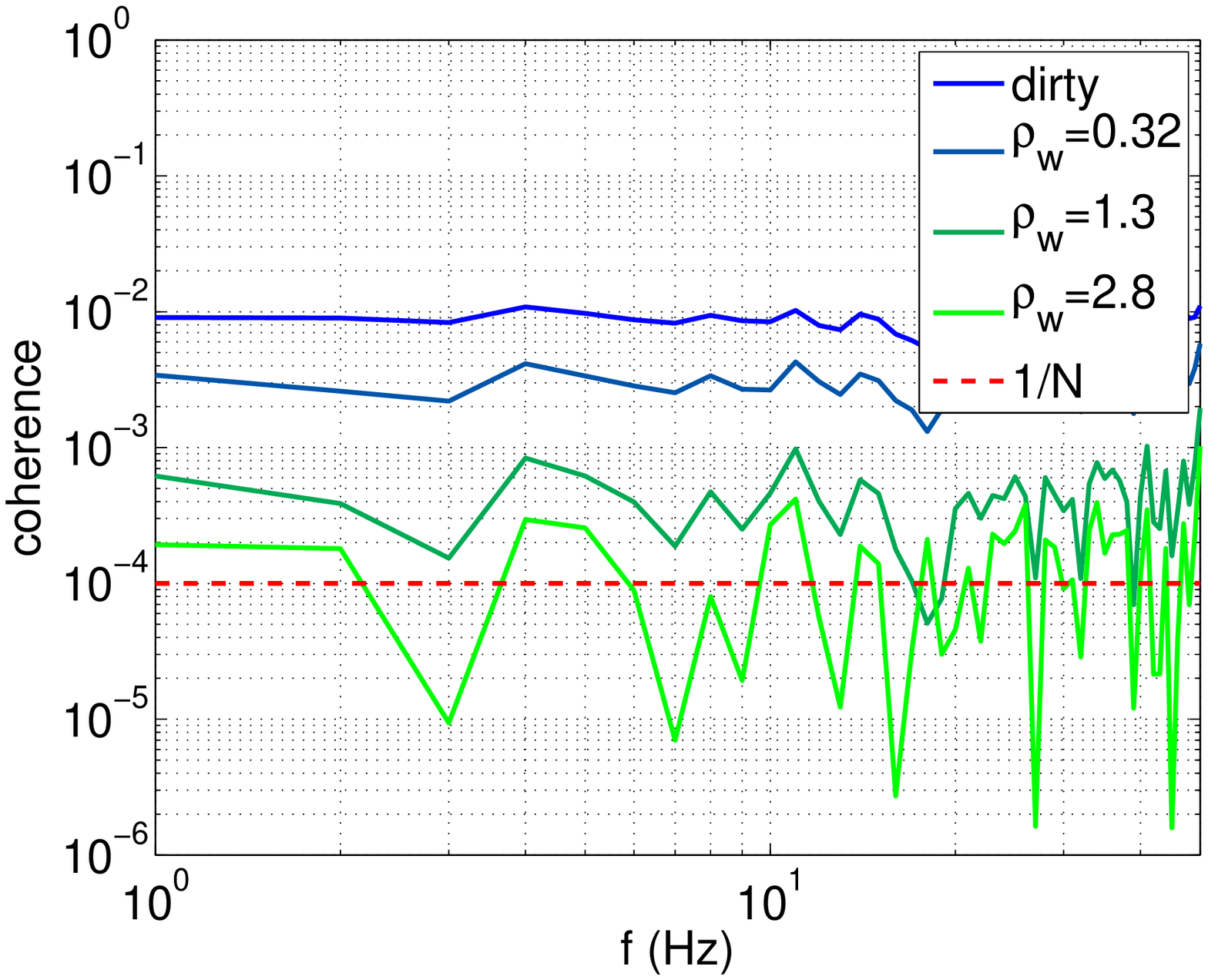, width=3.2in} &
    \psfig{file=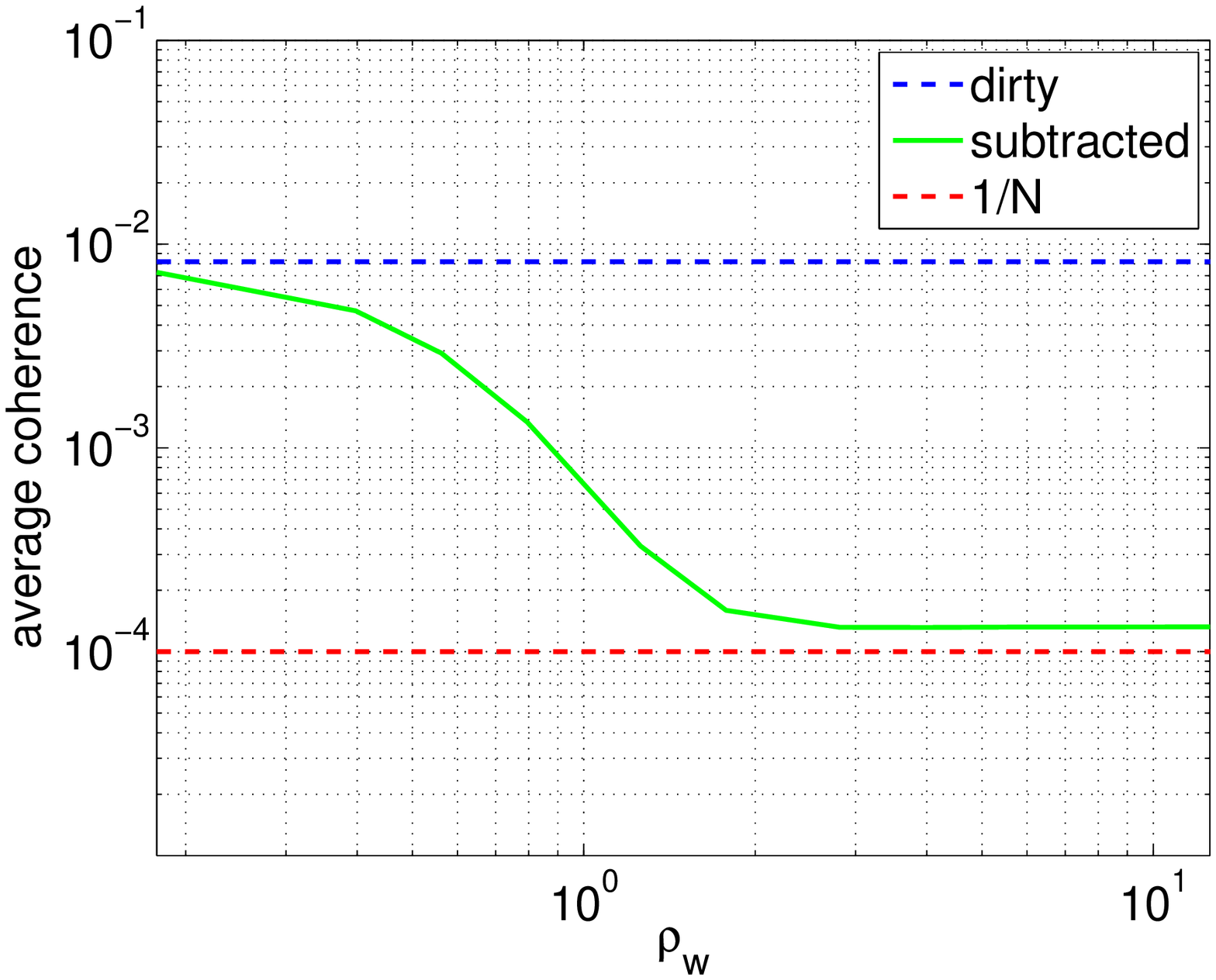, width=3.2in} \\
    \psfig{file=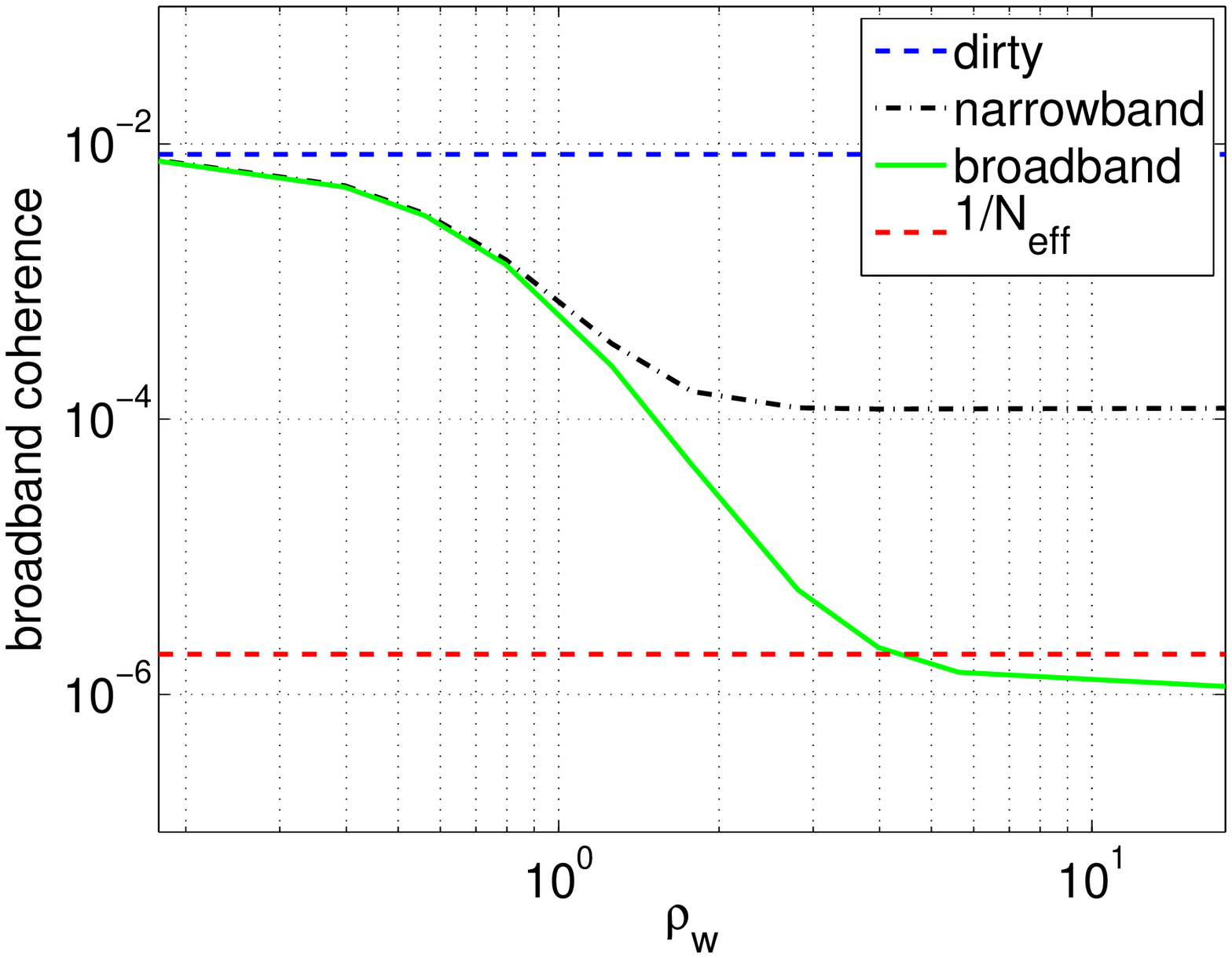, width=3.2in} &
    \psfig{file=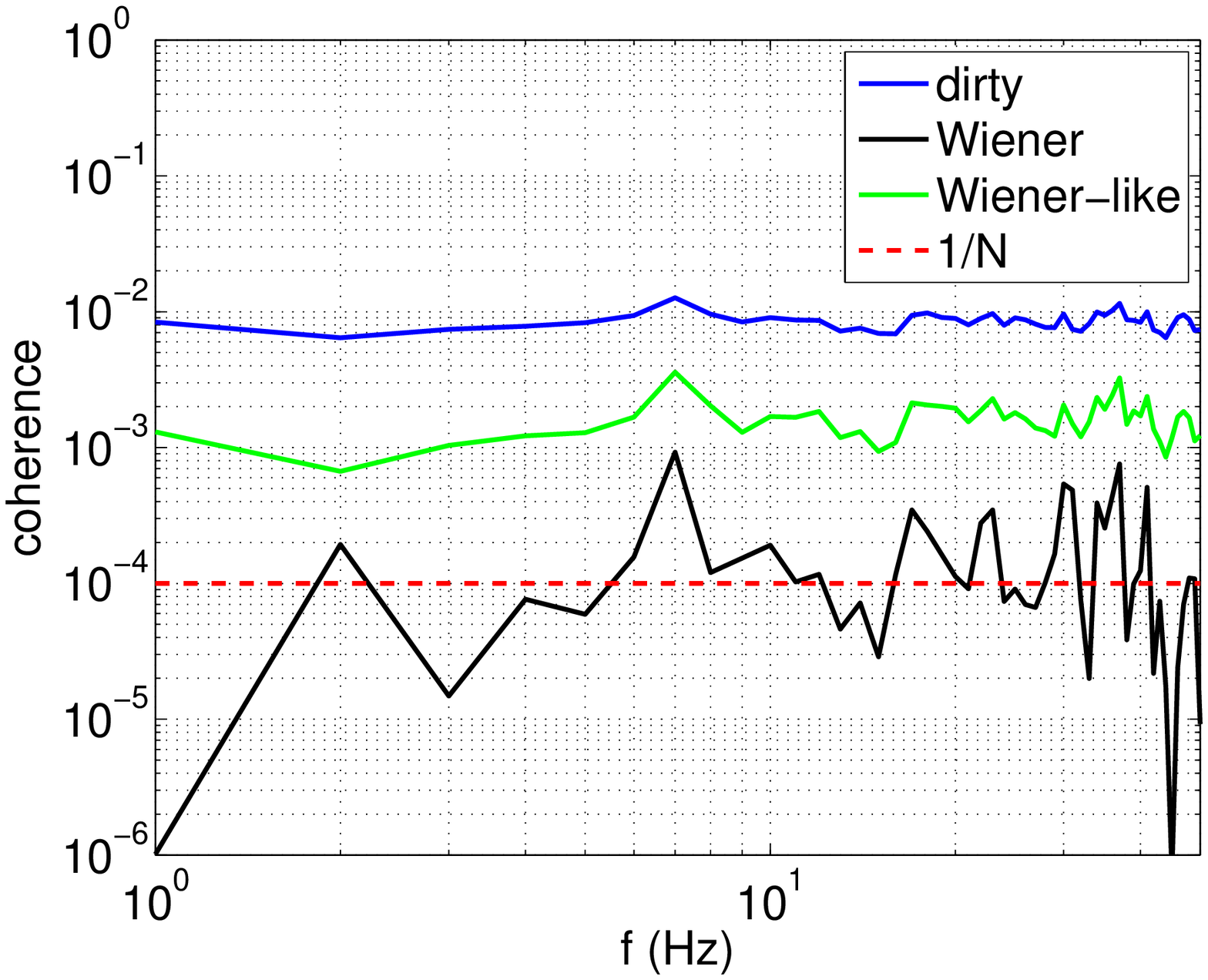, width=3.2in}
    % These plots are all made with src/rho_m.m
  \end{tabular}
  \caption{
    Reduced coherence with Wiener subtraction.
    Top-left: $\text{coh}(f)$ for different values of witness signal-to-noise ratio $\rho_w$; see Eq.~\ref{eq:rho_m}.
    The blue curve (dirty) represents no Wiener subtraction.
    The progressively greener curves indicate subtraction with progressively higher values of $\rho_w$.
    The dashed red $1/N$ curve indicates the expected level of coherence between two uncorrelated channels.
    Top-right: frequency-averaged coherence as a function of $\rho_w$.
    The dashed blue line indicates the coherence with no subtraction while the dashed red $1/N$ line indicates the expected level of coherence between two uncorrelated channels.
    Bottom-left: narrowband and broadband coherence (Eq.~\ref{eq:cohprime}) as a function of $\rho_w$.
    The dashed red line $1/N_\text{eff}$ indicates the expected level of broadband coherence between two uncorrelated channels.
    While $\rho_w\gtrsim2$ is sufficient to produce an apparently clean (narrowband) coherence spectrum, $\rho_w\gtrsim4$ is necessary to eliminate correlated noise at a level that is negligible as measured by the broadband coherence.
    Bottom-right: coherence spectrum comparing Wiener and Wiener-like filtering (Eq.~\ref{eq:wiener-like}) with $\rho_w=1.8$
    Blue indicates the dirty spectrum, black shows Wiener subtraction, and green is ``Wiener-like.''
%    All three curves fall on top of each other when $\rho_w\gtrsim5.5$.
    Both the Wiener curve and the Wiener-like curve fall on top of the $1/N$ line when $\rho_w\gtrsim5.5$.
  }
  \label{fig:rho_m}
\end{figure*}

The results are summarized in Fig.~\ref{fig:rho_m}.
Fig.~\ref{fig:rho_m}a shows $\text{coh}(f)$ for several different values of $\rho_w$.
(The frequency values shown on the abscissa are not important since our toy model uses white noise.)
When $\rho_w$ is large $\gtrsim2.8$, the coherence spectrum is consistent with $1/N$, which indicates that the cleaned data appear to be uncorrelated (up to one part in $N$).
For smaller values of $\rho_w$, however, there is clear excess coherence, indicating that the correlated noise is still detectable after Wiener filtering.

Fig.~\ref{fig:rho_m}b shows the frequency-averaged coherence from Fig.~\ref{fig:rho_m}a as a function of $\rho_w$.
The function has a sigmoid-like shape.
At small values of $\rho_w\lesssim0.2$, the subtraction is largely ineffective.
At high values of $\rho_w\gtrsim2$, the subtraction is largely effective at a level approaching the measurement uncertainty.

To this point, we have focused on the visibility of coherent structure in spectra.
However, searches for the stochastic background are sensitive to low-level {\it broadband} coherence beneath the dashed red $1/N$ lines in Fig.~\ref{fig:rho_m}.
By combining measurements from many frequency bins, stochastic searches gain sensitivity to both astrophysical signals {\it and} correlated noise; for a related discussion, see~\cite{locus}.
In order to demonstrate the importance of this effect, we define
\begin{equation}\label{eq:cohprime}
  \begin{split}
    \text{broadband coh} \equiv
    \frac{\left|\overline{\overline{\tilde{s}_1^*(f) \tilde{s}_2(f)}}\right|^2}
         {\overline{\overline{\left|\tilde{s}_1(f)\right|^2}} \,
           \overline{\overline{\left|\tilde{s}_2(f)\right|^2}} } \quad \text{(dirty)} \\
    \text{broadband coh} \equiv
    \frac{\left|\overline{\overline{\tilde{s}_1^{'*}(f) \tilde{s}'_2(f)}}\right|^2}
         {\overline{\overline{\left|\tilde{s}'_1(f)\right|^2}} \,
           \overline{\overline{\left|\tilde{s}'_2(f)\right|^2}} } \quad \text{(clean)} \\
  \end{split}
\end{equation}
$\text{Broadband coh}$ is analogous to $\text{coh}(f)$ except the double-overlines indicate averaging over both time {\it and} frequency.
Since our toy-model problem assumes white noise, we can use $\text{broadband coh}$ to investigate the efficacy of Wiener subtraction down to $1/N_\text{eff}$ where $N_\text{eff}=N n_\text{bins}$ and $n_\text{bins}$ is the number of frequency bins in each discrete Fourier transform.

In Fig.~\ref{fig:rho_m}c, we plot both narrowband (black) and broadband (green) coherence as a function of $\rho_w$.
When $\rho_w\gtrsim2$, the narrowband coherence spectra appears clean.
Examining the broadband coherence, however, it is apparent that there is still detectable correlated noise until $\rho_w\gtrsim4$.

\subsection{Wiener-like subtraction}
In this subsection, we consider an alternative approach to subtraction, which is similar in spirit to Wiener filtering, but slightly different.
Instead of carrying out Wiener subtraction as defined in Eq.~\ref{eq:wiener}, we subtract correlated noise as follows:
\begin{equation}\label{eq:wiener-like}
  \widehat{Y}_\text{WL}(f) \equiv \widehat{Y}(f) - 
  k\, \widehat{r}_1^* \widehat{r}_2 
  \left(\overline{\tilde{w}_1^*(f)\tilde{w}_2(f)}\right) .
\end{equation}
That is, instead of subtracting filtered witness channels from each dirty channel separately, the witness channels are cross-correlated and used to derive an estimate for the correlated strain noise power $H_M(f)$ (Eq.~\ref{eq:HM}).
The estimators for the filter functions $\widehat{r}_1$ and $\widehat{r}_2$ are calculated as before using Eq.~\ref{eq:r1r2}.
We call this approach ``Wiener-like'' filtering.

How does the Wiener-like subtraction compare to actual Wiener subtraction?
The answer to this question is illustrated in Fig.~\ref{fig:rho_m}d.
True Wiener filtering performs better than Wiener-like filtering, eliminating more correlated noise at a fixed value of $\rho_w$.
However, they both produce effective subtraction in the limit of $\rho_w\rightarrow\infty$.

\subsection{A priori subtraction}
Let us now imagine that the transfer functions are known with high precision.
For example, perhaps it is possible to generate a temporary excitation in order to measure the transfer functions with a high signal-to-noise ratio.
In this hypothetical scenario, Eq.~\ref{eq:wiener-like} becomes
\begin{equation}\label{eq:apriori}
   \widehat{Y}''(f) \equiv \widehat{Y}(f) -
   k\, r_1^* r_2 \left(\overline{\tilde{w}_1^*(f)\tilde{w}_2(f)}\right) .
\end{equation}
Note, we have removed the hats from the transfer functions and denoted the noise-subtracted power estimator with a double-prime.

\begin{figure*}[hbtp!]
  \begin{tabular}{cc}
    \psfig{file=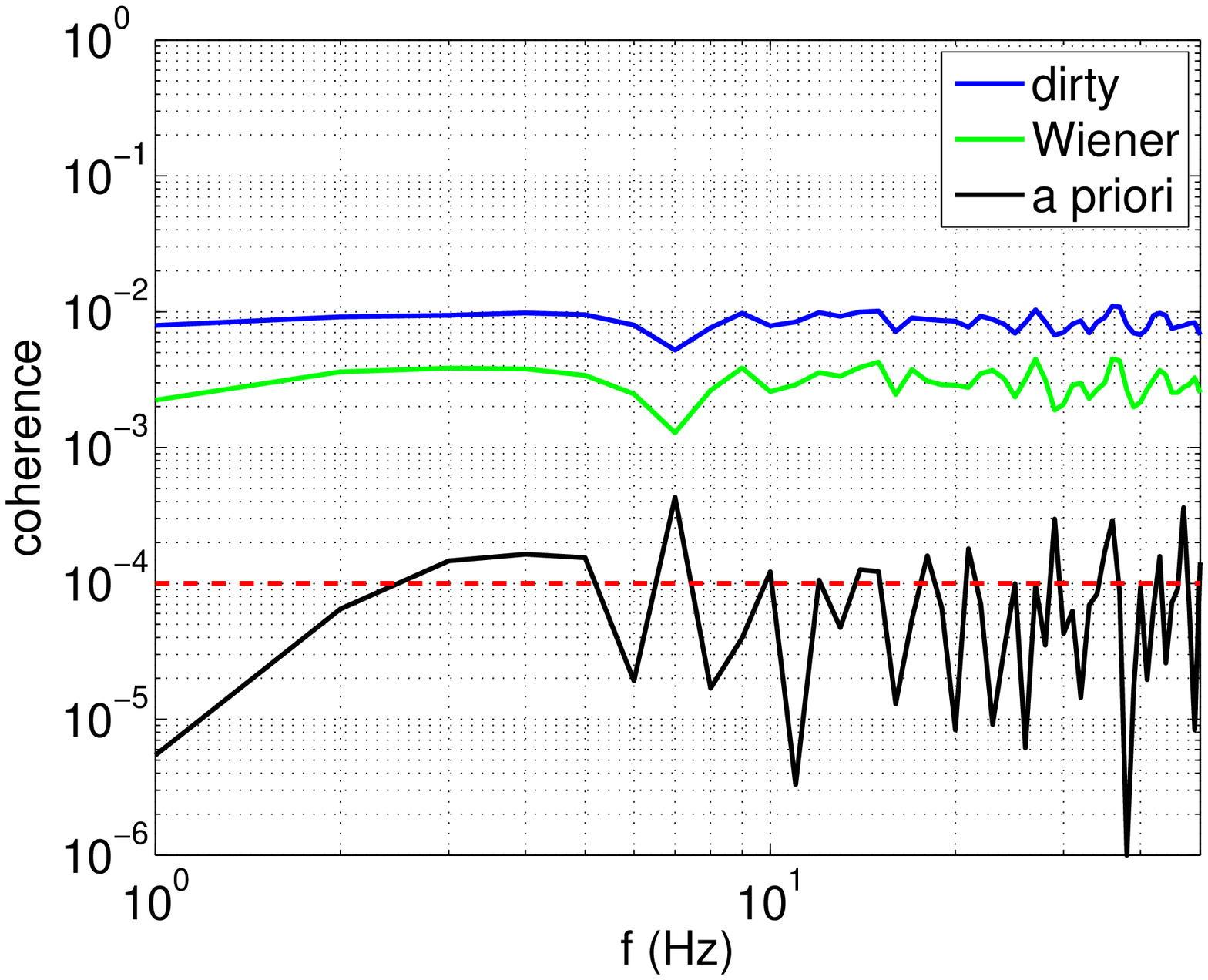, width=3.2in} &
    \psfig{file=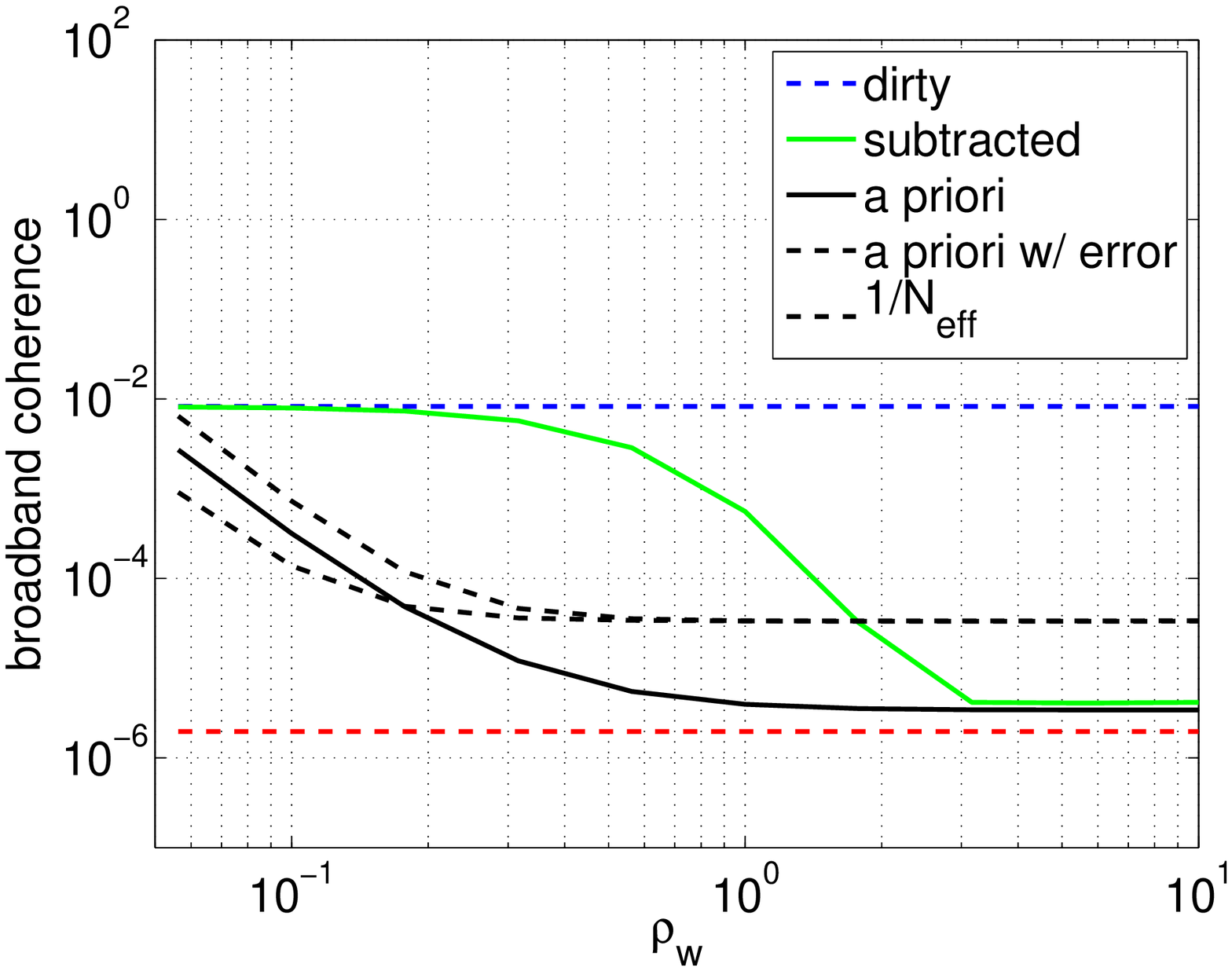, width=3.2in}
    % rho_m5.eps is made with src/rho_m.m and rho_m6.m is made with
    % apriori_bro2.m, which, in turn, is filled with information from rho_m.m.
  \end{tabular}
  \caption{
    Wiener-like subtraction when the transfer function is known a priori.
    Left: coherence spectrum for a pair of channels contaminated by correlated noise (blue), for the same channels cleaned with Wiener filtering (green), and cleaned with Wiener-like subtraction using the true transfer function; see Eq.~\ref{eq:apriori}.
    For this example, $\rho_w=0.56$.
    The dashed red line indicates the expected level of accidental correlation from two uncorrelated channels.
    Right: broadband coherence as a function of $\rho_w$.
    The dashed blue line indicates the value with no cleaning.
    The green line shows the improvement through Wiener subtraction.
    The solid black line demonstrates the improvement obtained using a priori subtraction with a precisely known transfer function.
    The dashed black lines show how this improvement is compromised when we apply a priori subtraction using a transfer function with a $\pm25\%$ amplitude error.
  }
  \label{fig:apriori}
\end{figure*}

How does Wiener-like subtraction perform given perfect knowledge of the transfer functions?
(We henceforth refer to this technique as ``a priori subtraction.'')
This question is answered by reference to Fig.~\ref{fig:apriori}.
The left-hand panel shows how a priori subtraction outperforms standard Wiener filtering at a fixed $\rho_w$.
The right-hand panel shows broadband coherence (Eq.~\ref{eq:cohprime}) as a function of $\rho_w$ for Wiener filtering and a priori filtering.
We observe that a priori filtering can achieve significant suppression of correlated noise for $\rho_w\lesssim1$ whereas Wiener filtering is only marginally effective in this regime.

In reality, the transfer function can only be known with finite precision.
Thus, it is interesting to ask: how does a priori subtraction perform when there is an error associated with $r$?
To answer this question, we examine the dashed black lines in Fig.~\ref{fig:apriori}b, which show the broadband coherence for a priori subtraction assuming a $\pm25\%$ error in the amplitude of the transfer function.
We observe that, over some range of values of $\rho_w$, a priori subtraction (with an error) still improves over Wiener filtering.
However, the broadband coherence flattens out at $\rho_w\gtrsim0.1$ ($\text{coh}\approx10^{-3}$).
For $\rho_w\gtrsim1$, Wiener filtering outperforms a priori subtraction with a 25\% amplitude error (see Fig.~\ref{fig:rho_m}).
Thus, a priori subtraction with a non-negligible error can perform both better and worse than Wiener filtering---it depends on the size of the transfer function error.

\subsection{Sectioned data}
% http://www.ldas-sw.ligo.caltech.edu/ilog/pub/ilog.cgi?group=stochastic&date_to_view=01/22/2013&anchor_to_scroll_to=2013:02:11:07:53:35-ethrane
In this subsection, we consider the question: how does the performance of Wiener filter subtraction change if the data are divided into sections to estimate the transfer functions $\widehat{r}_1$ and $\widehat{r}_2$ (see Eq.~\ref{eq:r1r2}).
In the toy-model calculation presented in the previous subsection, we calculated $\widehat{r}_1$ and $\widehat{r}_2$ using the entire dataset (consisting of $N_\text{segs}=10^4$, $\unit[1]{s}$-long segments sampled at $\unit[100]{Hz}$).
In doing so, we implicitly assume that the true transfer functions $r_1$ and $r_2$ are approximately stationary.
In fact, this need not be the case.
For example, the transfer function for magnetic noise coupling to angular degrees of freedom may exhibit non-stationarity due to the drifting alignment of the beam~\footnote{It may be possible account for slow alignment drift with a separate set of witness sensors devoted to monitoring the beam position.}.
However, non-stationarity can be taken into account by breaking the data into $N_\text{sections}$ sections over which the transfer function is approximately stationary.
Then, we can calculate the Wiener filter separately for each section.

Now we repeat the calculations from the previous subsection using a dataset, which is broken into sections, but which is otherwise identical.
We add a simulated monochromatic astrophysical signal $H(f)\propto\delta(f-f_0)$ in order to determine how the sensitivity to astrophysical signals changes with $N_\text{sections}$.
(The astrophysical signal appears in both $s_1$ and $s_2$, but not in the witness channels $w_1$ and $w_2$.)
We assume a high witness signal-to-noise ratio $\rho_w=5.6$.

\begin{figure}[hbtp!]
  \begin{tabular}{c}
    \psfig{file=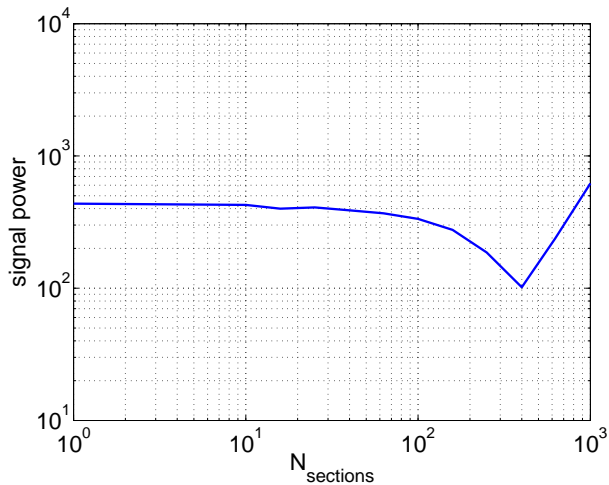, width=3.2in}
    % calculated w rho_test_harvest3 using rho_w = sqrt(sqrt(0.1) * 100) = 5.6
  \end{tabular}
  \caption{
    Apparent signal power as a function of the number of data sections $N_\text{sections}$ used to calculate the Wiener filters.
    Between $N=1$--$400$, the observed power gradually (and undesirably) falls with increasing $N_\text{sections}$ due to a systematic error from Wiener filtering.
    The turning point at $N_\text{sections}=400$ is due to the growth of broadband noise injected into the strain channels through the Wiener filters.
    (This is also an undesirable effect and does not imply improved measurement of signal power, only an increase in noise.)
    We use a witness signal-to-noise ratio of $\rho_w=5.6$.
  }
  \label{fig:nsegs}
\end{figure}

The results are summarized in Fig.~\ref{fig:nsegs}, which plots the signal power evaluated at $f_0$ (the frequency of the astrophysical signal) as a function of $N_\text{sections}$.
%Larger values of $\text{coh}(f_0)$ imply a more detectable astrophysical signal than smaller values.
As $N_\text{sections}$ is increased, the signal power begins to fall.
This is due to the fact that the astrophysical signal power is reduced by Wiener filtering.
Then, around $N_\text{sections}=400$, the signal power appears to increase.
However, this power increase occurs across the entire band (at frequencies where no signal is present).
This is because Wiener filtering, in this regime, injects significant noise from the witness sensors into the strain channels.
The loss of signal power and the injection of witness noise are both undesirable.
Thus, it is preferable to use as few sections of data as possible.
However, for the example shown here, we note that there is a range of values of $N_\text{sections}$ for which the loss of signal power is relatively modest.

\subsection{Instrumental noise vs local environmental noise}\label{epsilon}
In Eqs.~\ref{eq:goes_to_r}--\ref{eq:goes_to_zero}, we investigated the high-$\rho_w$ and low-$\rho_w$ behavior of the transfer function estimator obtained through Wiener filtering.
In our derivation, we assumed that witness sensor channel $w_I$ could be written as a sum of instrumental noise $\eta_I$, (which does not couple to the strain channel $s_I$) and an environmental noise $m$, (which does couple to $s_I$).
However, it is possible to consider a more general version of Eq.~\ref{eq:w12}:
\begin{equation}\label{eq:w12p}
  \begin{split}
    w'_1(t) & = m(t) + m_1(t) + \eta_1(t) \\
    w'_2(t) & = m(t) + m_2(t) + \eta_2(t) .
  \end{split}
\end{equation}
Eq.~\ref{eq:w12p} includes {\it instrumental noise} $\eta_1(t)$ and $\eta_2(t)$, defined as the sum of all witness noise, which does not couple to the strain channels, e.g., electronic noise.
However, we add additional {\it local environmental noise} $m_1(t)$ and $m_2(t)$, which couples into $s_1$ and $s_2$ with transfer functions $\mathfrak{r}_1\neq r_1$ and $\mathfrak{r}_2\neq r_2$ respectively:
\begin{equation}
  \begin{split}
    s_1 & = h_1 + n_1 + r_1 m + \mathfrak{r}_1 m_1 \\
    s_2 & = h_2 + n_2 + r_2 m+ \mathfrak{r}_2 m_2 .
  \end{split}
\end{equation}

We define the local environmental noise power spectral densities (assumed to be the same at each witness sensor) to be:
\begin{equation}
  \begin{split}
    {\cal M}(f) = k\, \langle\left|\tilde{m}_1(f)\right|^2\rangle = k\, 
    \langle\left|\tilde{m}_2(f)\right|^2\rangle .
  \end{split}
\end{equation}
[Note that ${\cal M}(f)$ is local witness noise power spectrum whereas $M(f)$ is the correlated magnetic field noise.]
In the case of magnetic noise, $\mathfrak{r}_1$ represents the $w$-to-$s$ transfer function of locally generated noise in the vicinity of a single test mass whereas $r_1$ represents the transfer function of Schumann resonance fields, which occur over a wide area and with a different geometry than the locally generated fields.
We assume that $\langle\tilde{m}_1^*\tilde{m}_2\rangle=0$ and $\langle\tilde{m}_1^*\tilde{m}\rangle=\langle\tilde{m}_2^*\tilde{m}\rangle=0$.
We further assume that $\mathfrak{r}_1=\mathfrak{r}_2\equiv\mathfrak{r}$.

In order to investigate the relative importance of local environmental noise versus instrumental noise, we introduce the following parameterization:
\begin{equation}\label{eq:epsilon}
  \begin{split}
    {\cal M}(f) = \epsilon\,\Pi(f) \\
    {\cal N}(f) = (1-\epsilon) \, \Pi(f)
  \end{split}
\end{equation}
so that the total noise (instrumental + local environmental) is fixed
\begin{equation}
  \Pi(f) \equiv {\cal M}(f) + {\cal N}(f) 
\end{equation}
The variable $\epsilon$ determines how much witness noise is local versus instrumental.
When $\epsilon=0$, the witness noise is entirely instrumental (and does not couple to the dirty channel) and when $\epsilon=1$, the witness noise is entirely local (and does couple to the witness channel).
It follows that 
\begin{equation}\label{eq:rho_w}
  \rho_w^2 = \frac{M(f)}{\Pi(f)} = \frac{M(f)}{{\cal M}(f) + {\cal N}(f)}
\end{equation}
does not depend on $\epsilon$.

\begin{figure*}[hbtp!]
  \begin{tabular}{cc}
    \psfig{file=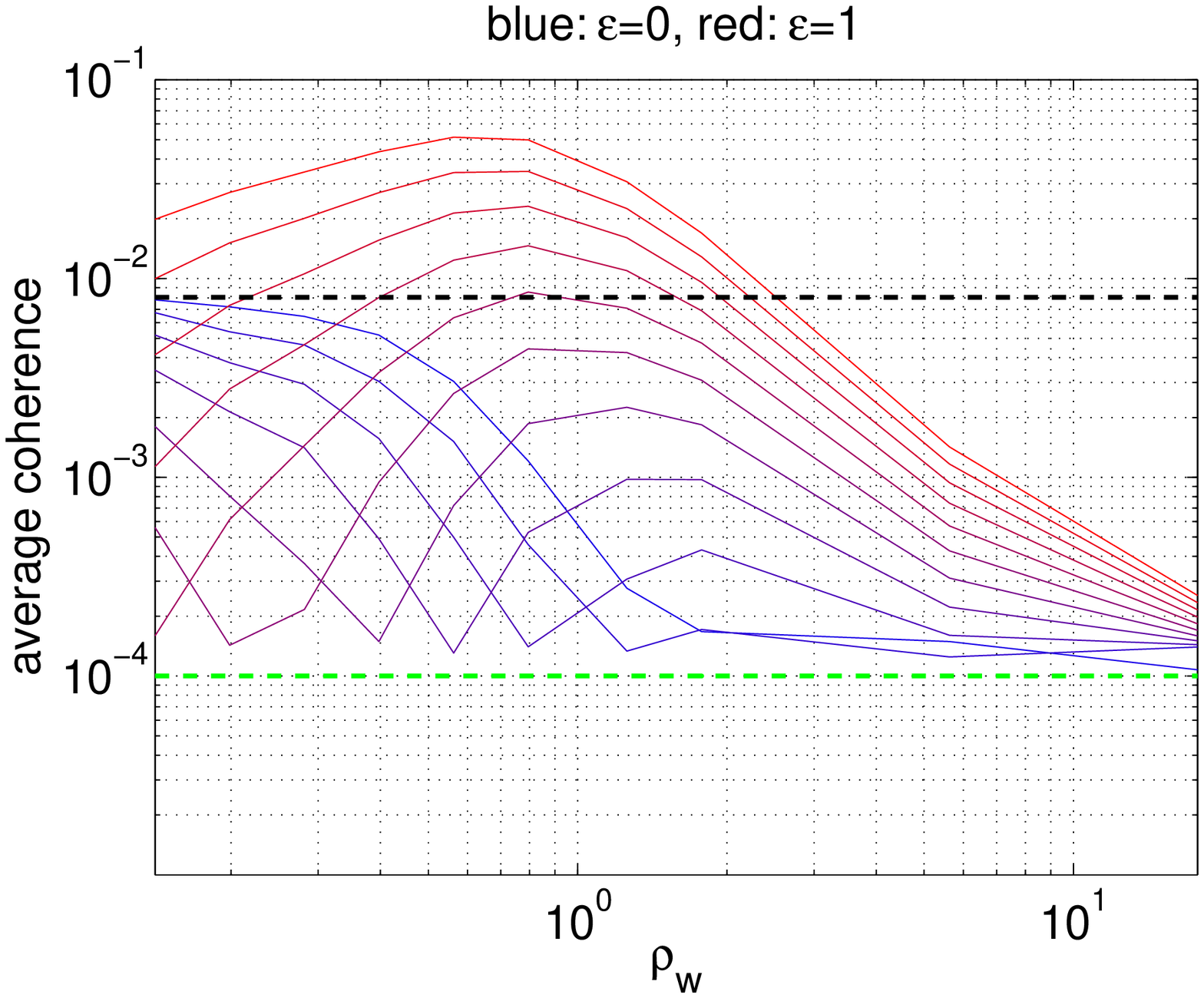, width=3.2in} &
    \psfig{file=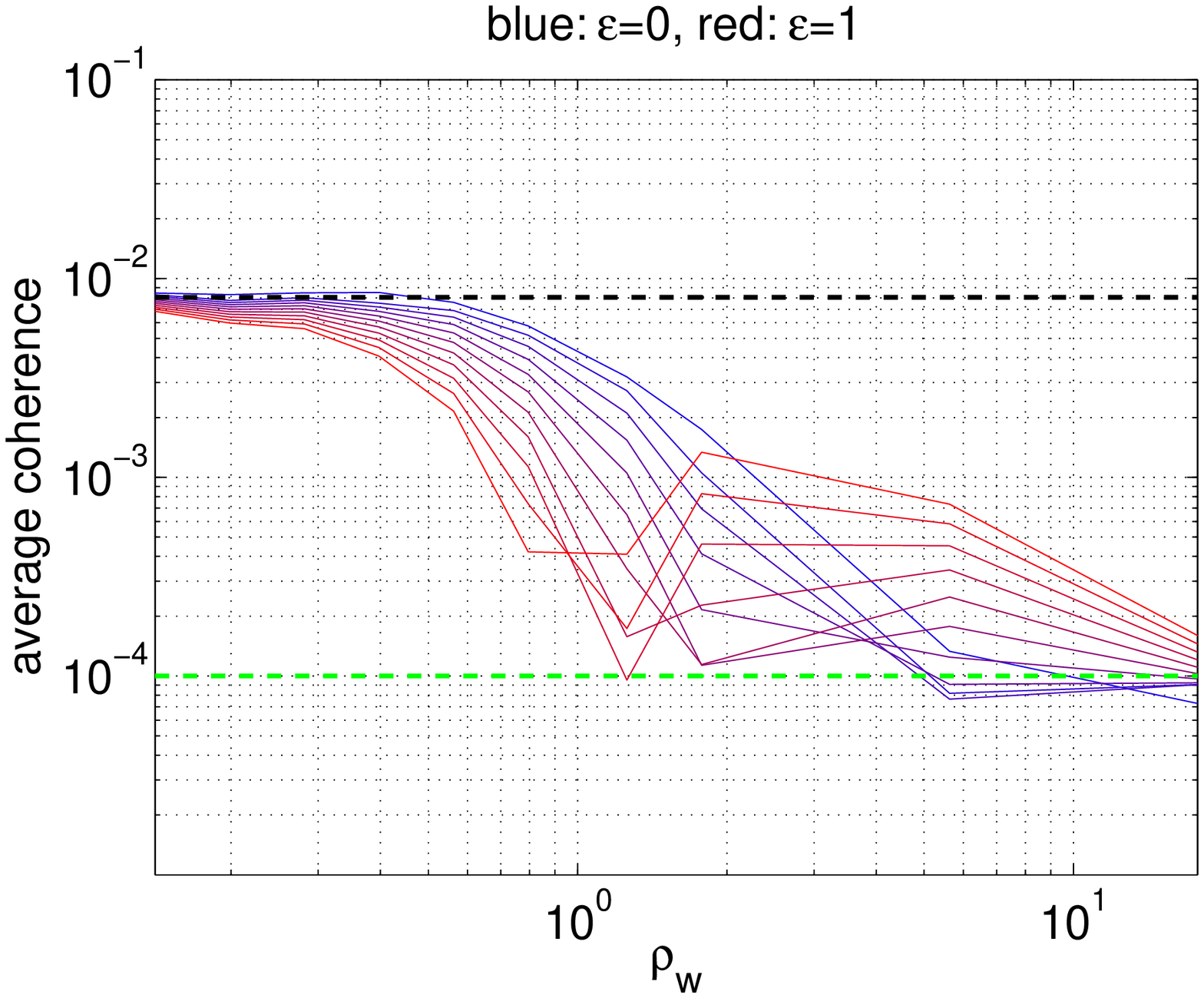, width=3.2in} \\
    \psfig{file=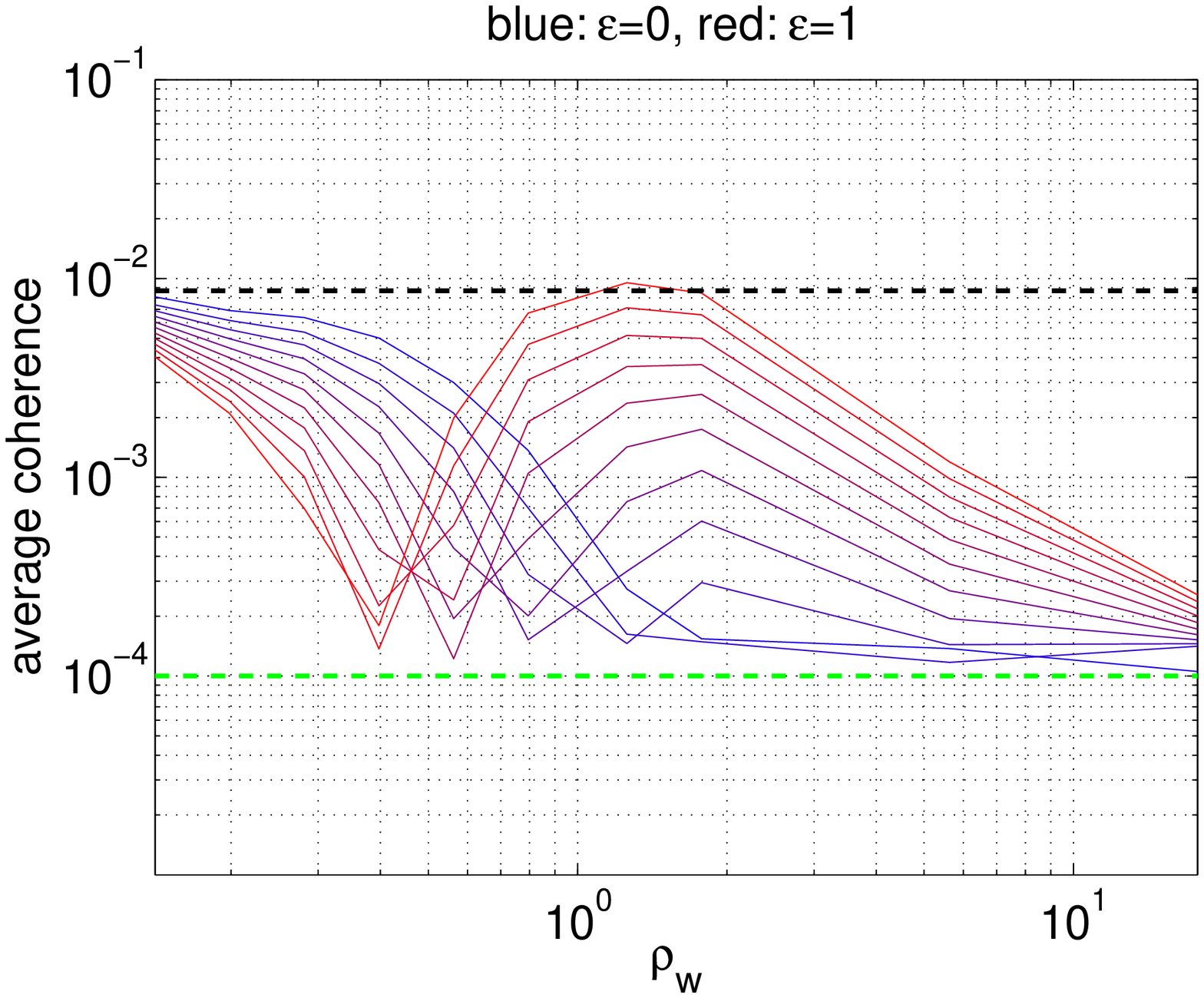, width=3.2in} &
    \psfig{file=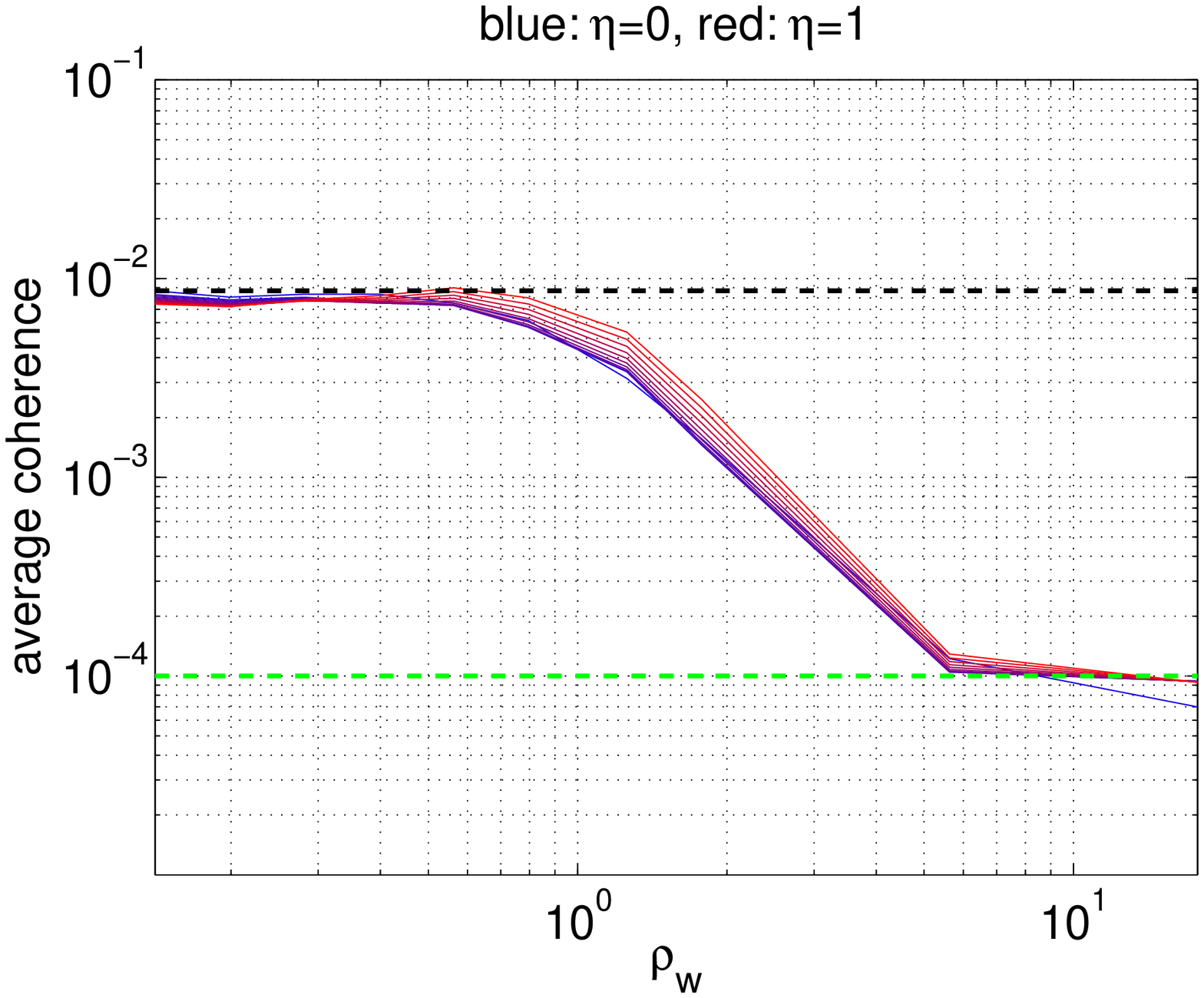, width=3.2in} \\
    % These plots are all made with src/localnoise.m and rho_w.m.
  \end{tabular}
  \caption{
    Coherence spectra for varying degrees of local environmental noise vs instrumental noise.
    The two rows correspond to a sign difference between the local environmental and instrumental transfer functions; see Eq.~\ref{eq:cases}.
    The top row is case one [$\text{phase}(r)=\text{phase}(\mathfrak{r})$] while the bottom row is case two [$\text{phase}(r)=\text{phase}(\mathfrak{r})+\pi$].
    The left columns shows Wiener subtraction while the right shows Wiener-like subtraction.
    In each panel, the dashed black line represents the coherence before subtraction and the dashed green line represents a perfectly clean spectrum.
    Each trace shows the coherence following subtraction.
    The colors show the results for different evenly spaced values of $\epsilon$ (Eq.~\ref{eq:epsilon}).
    The bluest hue corresponds to $\epsilon=0$ (pure instrumental noise) while the reddest hue corresponds to $\epsilon=1$ (pure local environmental noise).
  }
  \label{fig:local}
\end{figure*}

We vary $\epsilon$ for a fixed value of $\rho_w$ and determine how this affects subtraction.
We consider two special cases:
\begin{equation}\label{eq:cases}
  \begin{split}
    \text{phase}(r)=\text{phase}(\mathfrak{r}) & \quad \text{(case one)} \\
    \text{phase}(r)=\text{phase}(\mathfrak{r})+\pi & \quad \text{(case two)}
  \end{split}
\end{equation}
The results are summarized in Fig.~\ref{fig:local}.
The top and bottom rows show coherence spectra for cases one and two respectively.
The left column shows results for Wiener filtering while the right shows results for Wiener-like filtering.
The different shades correspond to different values of $\eta=0$ (blue), $\eta=1$ (red), and evenly-spaced intermediate values.

At first glance, the results are mysterious.
In Fig.~\ref{fig:local}a, for example, it appears that an $\epsilon=50\%$ mixture of local environmental and instrumental noise (purple) allows us to carry out effective subtraction at much lower values of $\rho_w$ than can be obtained with purely instrumental noise $\epsilon=0$ (blue).
However, this apparent improvement is illusory.
Since local environmental noise couples to the dirty channel, its presence biases our estimate of $r_1$ and $r_2$.
This bias combines with the existing bias described in Eqs.~\ref{eq:goes_to_r}--\ref{eq:goes_to_zero}, which arises from the fact that $r_1$ and $r_2$ are always underestimated for finite $\rho_w$.
For some values of $\epsilon$ and $\rho_w$, the two biases offset somewhat, leading to what might be called ``accidentally successful subtraction.''

However, the bias from $\epsilon>0$ can cause very undesirable effects.
It is possible to subtract too much, removing the correlated noise {\it and} some or all of the correlated signal power $H(f)$.
In the parlance of gravitational-wave data quality jargon, Wiener filtering with $\epsilon>0$ is not ``safe''~\cite{smith_glitch,stamp_glitch}.
To be clear, the concern is not that the Wiener filter will coherently subtract the gravitational signal---after all, the magnetometer has no sensitivity to gravitational waves---but rather, that the Wiener filtered data may include anti-correlated noise, which {\em incoherently} cancels part of the gravitational-wave signal power.
Also, when $\epsilon>0$, it is possible to make the correlated noise worse.
In sum, given our assumed uncertainty about $\rho_w$, $\epsilon$, and the phase and amplitude differences between $r$ and $\mathfrak{r}$, subtraction with a non-zero value of $\epsilon$ is less reliable and less effective than $\epsilon=0$ subtraction.
We therefore recommend using witness sensors with essentially no local environmental noise (that couples into $s_1$ and $s_2$).

For the case of Schumann subtraction, this may be achieved by placing magnetometers sufficiently far away from the gravitational-wave detectors.
Alternatively, one could use magnetometers at one detector as the witness sensor for the other detector and vice versa.
This would ensure that the local magnetic fields measured by each witness sensor do not couple to the strain channels, and so we could treat the local magnetic noise as instrumental.

\subsection{Multiple witness sensors}\label{multiple}
In this subsection, we investigate how the efficacy of Wiener filtering changes with the inclusion of additional witness sensors.
Let us suppose that there are $n_w$ sensors at each site.
The time series for the $j^\text{th}$ sensors can be written as a sum of correlated noise and witness noise:
\begin{equation}
  \begin{split}
    w_{1,j}(t) = m(t) + \eta_{1,j}(t) \\
    w_{2,j}(t) = m(t) + \eta_{2,j}(t)
  \end{split}
\end{equation}
Here, the subscript $1,j$ represents the $j^\text{th}$ sensor at site~$1$.

There are two interesting limiting cases to consider.
In the first case, the witness noise in every sensor is completely correlated:
\begin{equation}\label{eq:many_sensors}
  \begin{split}
    \eta_{1,j}(t) = \eta_{1,k}(t) & \quad \forall (j,k) \\
    \eta_{2,j}(t) = \eta_{2,k}(t) & \quad \forall (j,k)
  \end{split}
\end{equation}
This is what we would expect to happen, e.g., if we placed several magnetometers  (with the same orientation) in close proximity at a location where large local magnetic fields induce witness noise far above the electronic noise in each sensor.
In this case, each channel contains the exact same information as every other channel.
Thus, there is no advantage to be gained by combining them.

Of course, closely placed magnetometers with {\em perpendicular} orientations may contain complementary information.
As noted above, a realistic subtraction scheme should utilize sensors measuring the magnetic field in three orthogonal directions.
For the sake of simplicity, however, we proceed considering magnetic field measurements for a single direction.

In the second case, the witness noise in any two sensors is completely uncorrelated:
\begin{equation}
  \begin{split}
    \langle\tilde\eta_{1,j}^*\tilde\eta_{1,k}\rangle=0 & \quad \forall (j\neq k) \\
    \langle\tilde\eta_{2,j}^*\tilde\eta_{2,k}\rangle=0 & \quad \forall (j\neq k) .
  \end{split}
\end{equation}
This case may obtain if it is possible to place each sensor with sufficient separation so that the local environmental noise is different for each one.
In this case, we may straightforwardly construct effective witness channels $w^\text{eff}$ by averaging together the $n_w$ channels:
\begin{equation}
  \begin{split}
    w^\text{eff}_1(t) & \equiv \frac{1}{n_w} \sum_{j=1}^{n_w} w_{1,j}(t) =
    m(t) + \frac{1}{n_w} \sum_{j=1}^{n_w} \eta_{1,j}(t) \\
    w^\text{eff}_2(t) & \equiv \frac{1}{n_w} \sum_{j=1}^{n_w} w_{2,j}(t) =
    m(t) + \frac{1}{n_w} \sum_{j=1}^{n_w} \eta_{2,j}(t) .
  \end{split}
\end{equation}
The correlated noise from each witness channel adds coherently while the witness noise (by assumption) adds incoherently.
Thus, the witness signal-to-noise ratio is 
\begin{equation}
  \rho_w^2 = M(f) \Big/ \frac{1}{n_w^2} \sum_{j=1}^{n_w} \Pi_j(f) ,
\end{equation}
where $\Pi_j(f)$ is the witness noise in channel $j$.
If we assume that every sensor has the same witness noise power spectrum $\Pi(f)$, we obtain
\begin{equation}
  \rho_w^2 = n_w \frac{M(f)}{\Pi(f)} .
\end{equation}
Thus, if it is possible to utilize witness sensors with independent witness noise, one can boost the witness signal-to-noise ratio by a factor of $n_w^{1/2}$ compared to just one witness sensor.

In between these two limiting cases, it is possible to have a set partially correlated channels.
In this case, one can construct an effective witness channel following the formalism from~\cite{ligo_wiener}.
However, the enhancement in $\rho_w$ is limited to a factor of $\leq n_w^{1/2}$.

In the following section, we show that it may be necessary to achieve a witness noise of $\Pi^{1/2}\lesssim\unit[0.2]{pT\,Hz^{-1/2}}$ (in the relevant $10$--$\unit[60]{Hz}$ band) in order to achieve suitable subtraction in one year of Advanced LIGO data.
Previously reported witness noise from local magnetic fields in this band $\Pi^{1/2}\approx4$--$\unit[20]{pT\,Hz^{-1/2}}$ is higher by a factor of $16$--$83\times$~\cite{schumann_ligo}.
If we imagine constructing an array of widely spaced magnetometers with comparable witness noise {\it assumed to be uncorrelated between sensors}, it could require $280$--$7000$ sensors to achieve an effective noise level of $\Pi^{1/2}\lesssim\unit[0.2]{pT\,Hz^{-1/2}}$ necessary for successful subtraction.
On the other hand, if we assume that the witness noise in each magnetometer can be reduced to  $\Pi^{1/2}\approx\unit[1]{pT\,Hz^{-1/2}}$ by moving them away from anthropogenic sources, then it may be sufficient to employ an array of $\approx$$18$ magnetometers.

\begin{table}
  \begin{tabular}{|c|c|}
    \hline {\bf power spectra} & {\bf meaning} \\\hline
    $H=k\,\langle\tilde{h}^*_1\tilde{h}_2\rangle$ & astrophysical strain \\\hline
    $H_M=k\,\langle r^*r \, \tilde{m}^*_1 \tilde{m}_2\rangle$ & correlated strain noise \\\hline
    $M=k\,\langle \tilde{m}^*_1 \tilde{m}_2\rangle$ & correlated magnetic field noise \\\hline
    $P_I=k\,\langle \tilde{s}_I^* \tilde{s}_I \rangle$ & detector $I$ strain auto-power \\\hline
    ${\cal N}_I=k\,\langle \tilde\eta_I^*\tilde\eta_I \rangle$ & sensor $I$ {\it instrumental} witness noise \\\hline
    ${\cal M}_I=k\,\langle \tilde{m}_I^*\tilde{m}_I \rangle$ & sensor $I$ {\it local} environmental witness noise \\\hline
    $\Pi_I={\cal M}_I + {\cal N}_I$ & sensor $I$ {\it total} witness noise \\\hline
  \end{tabular}
  \caption{Definitions of power spectra.}
  \label{tab:power}
\end{table}

\subsection{Take-away}
There are a number of important take-away messages from this section, which we enumerate here.
\begin{enumerate}
  \item The degree to which correlated noise can be successfully removed using Wiener (and Wiener-like) filtering is determined almost entirely by the witness signal-to-noise ratio $\rho_w$ with which the witness sensors can measure the correlated noise.
  \item Wiener filtering is largely limited by imperfect knowledge of the transfer function between the witness sensors and the dirty channels.
    If the transfer function can be measured independently with a high degree of precision, the required $\rho_w$ for successful subtraction is dramatically reduced.
  \item Coherence spectra can be deceptive: statistically important broadband correlated noise can lurk underneath a seemingly clean-looking coherence spectrum.
    It is therefore important to use a broadband statistic when evaluating the efficacy of subtraction in the context of a search for a broadband astrophysical signal.
  \item All else equal, instrumental noise (which does not couple to the dirty channels) is preferential to local environmental noise (which does couple to the dirty channels).
    Local environmental noise can lead to accidental (incoherent) subtraction of astrophysical signal power or exacerbated correlated noise.
  \item It is helpful to employ additional (perpendicular triplets of) witness sensors only if the witness noise in different sensors is uncorrelated.  In this case, $\rho_w$ increases like the square root of the number of sensors.
\end{enumerate}

\section{Prospects for subtraction in advanced detector networks}\label{aligo}
\subsection{Numerical simulation}
In order to assess the prospects for the subtraction of correlated noise in the Advanced LIGO network, we employ a numerical model.
While we focus here on Advanced LIGO, we expect our results to be representative of an arbitrary global network since Schumann fields are correlated over global distances~\cite{schumann_ligo}.

For each detector, we simulate uncorrelated Gaussian strain noise $n_1$ and $n_2$ using the design sensitivity curve for Advanced LIGO~\cite{noisecurve}.
Next, we simulate correlated Gaussian magnetic field noise $m$ using a previously measured power spectrum~\cite{schumann_ligo}.
We simulate purely instrumental noise in the witness sensors $\eta_1$ and $\eta_2$ ($\epsilon=0$; see Subsection~\ref{epsilon}).
The sensor noise is simulated assuming a white noise spectrum, which is consistent with previous measurements~\cite{schumann_ligo}.
Previously reported witness noise, which is primarily due to local magnetic fields, is $\Pi^{1/2}\approx4$--$\unit[20]{pT\,Hz^{-1/2}}$~\cite{schumann_ligo}.
% see: https://ldas-jobs.ligo.caltech.edu/~ethrane/files/tfcoh/robert/S5_H_BSC5_MAGZ_vs_L_EX_MAGZ_power.png
We treat $\Pi^{1/2}$ as a tunable parameter.
By varying $\Pi^{1/2}$, with a fixed correlated noise power spectrum ${\cal M}(f)$, we determine $\rho_w$; see Eq.~\ref{eq:rho_w}.

\begin{figure*}[hbtp!]
  \begin{tabular}{cc}
    \psfig{file=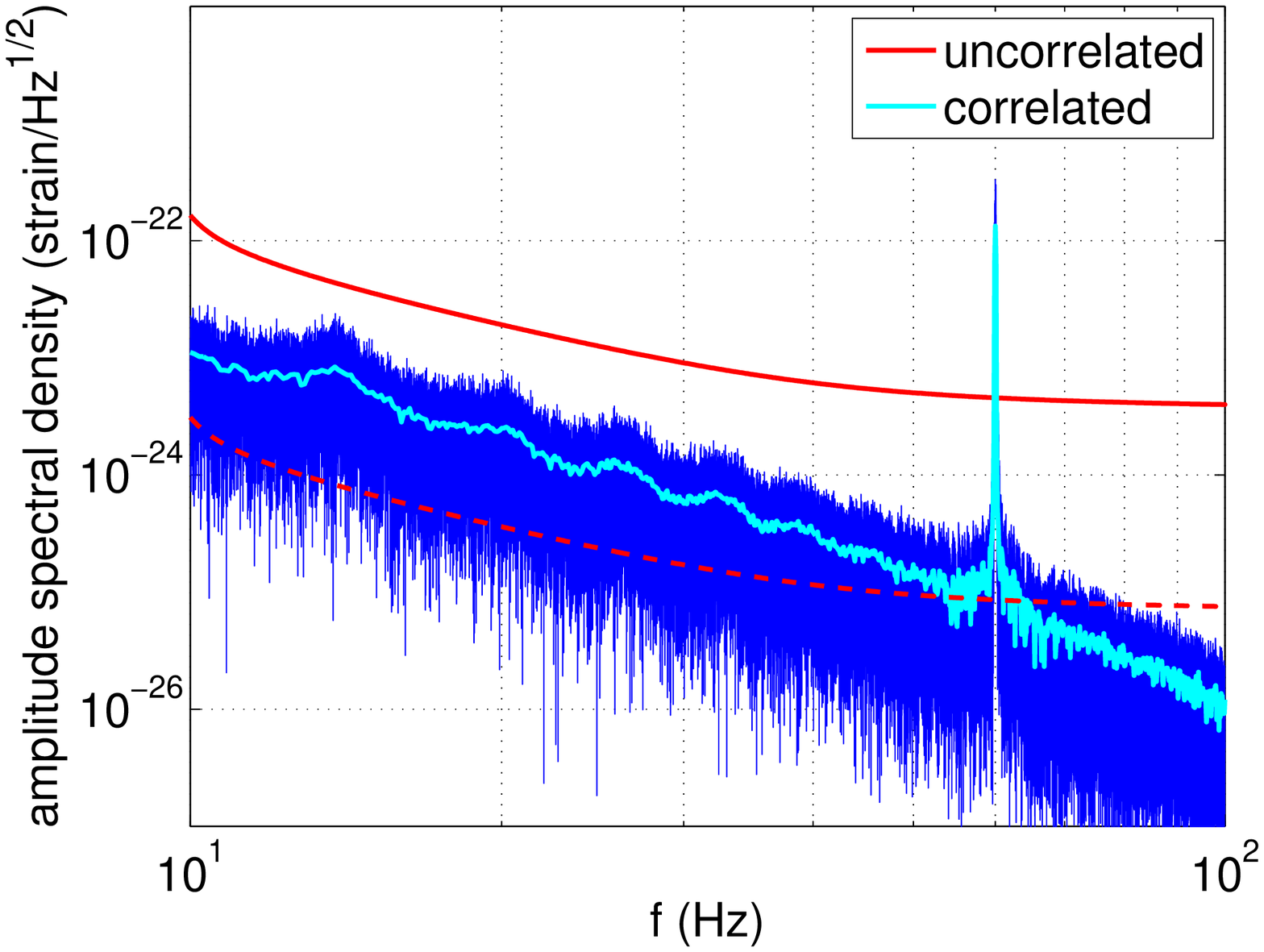, width=3.2in} & 
    \psfig{file=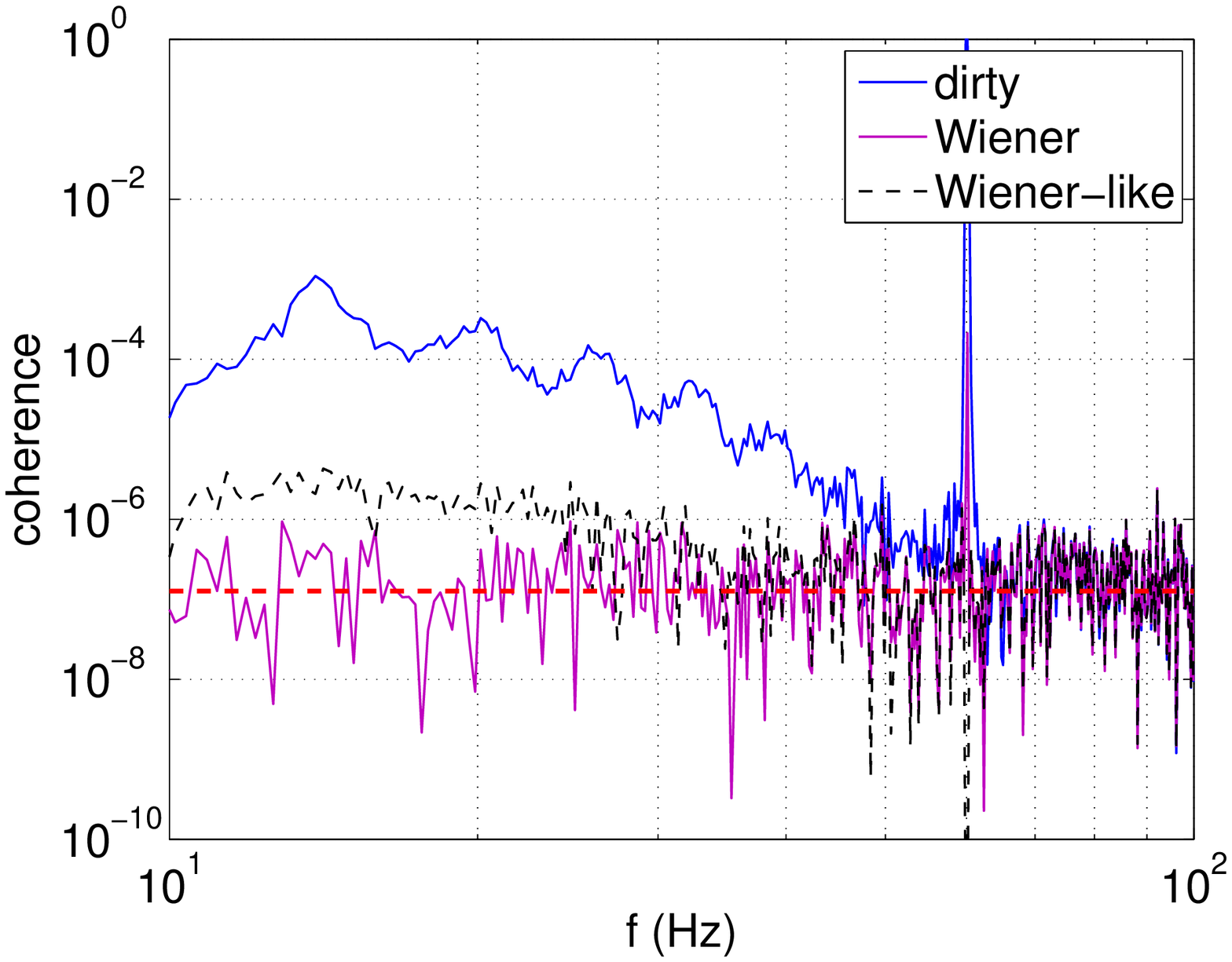, width=3.2in} \\
    \psfig{file=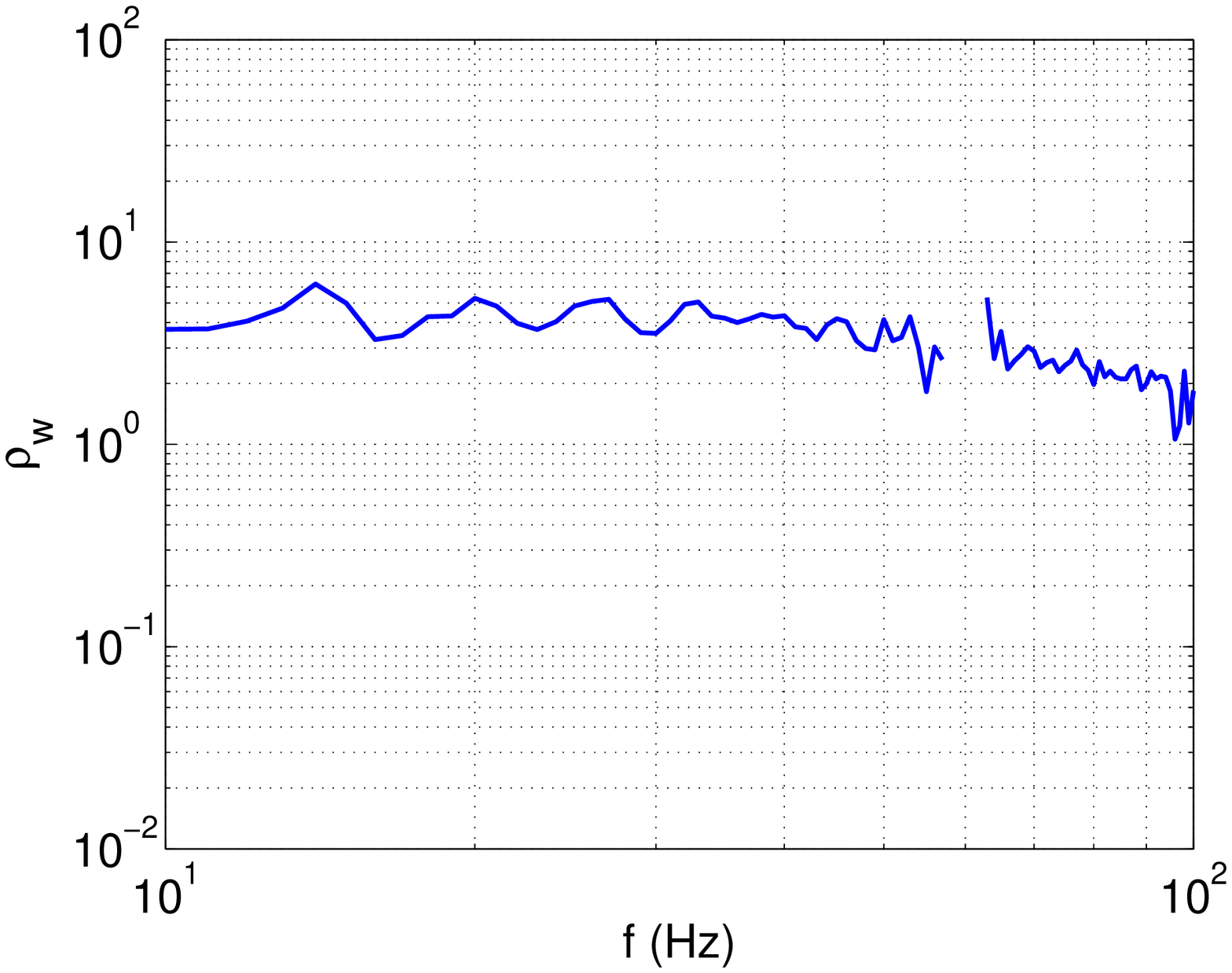, width=3.2in} &
    \psfig{file=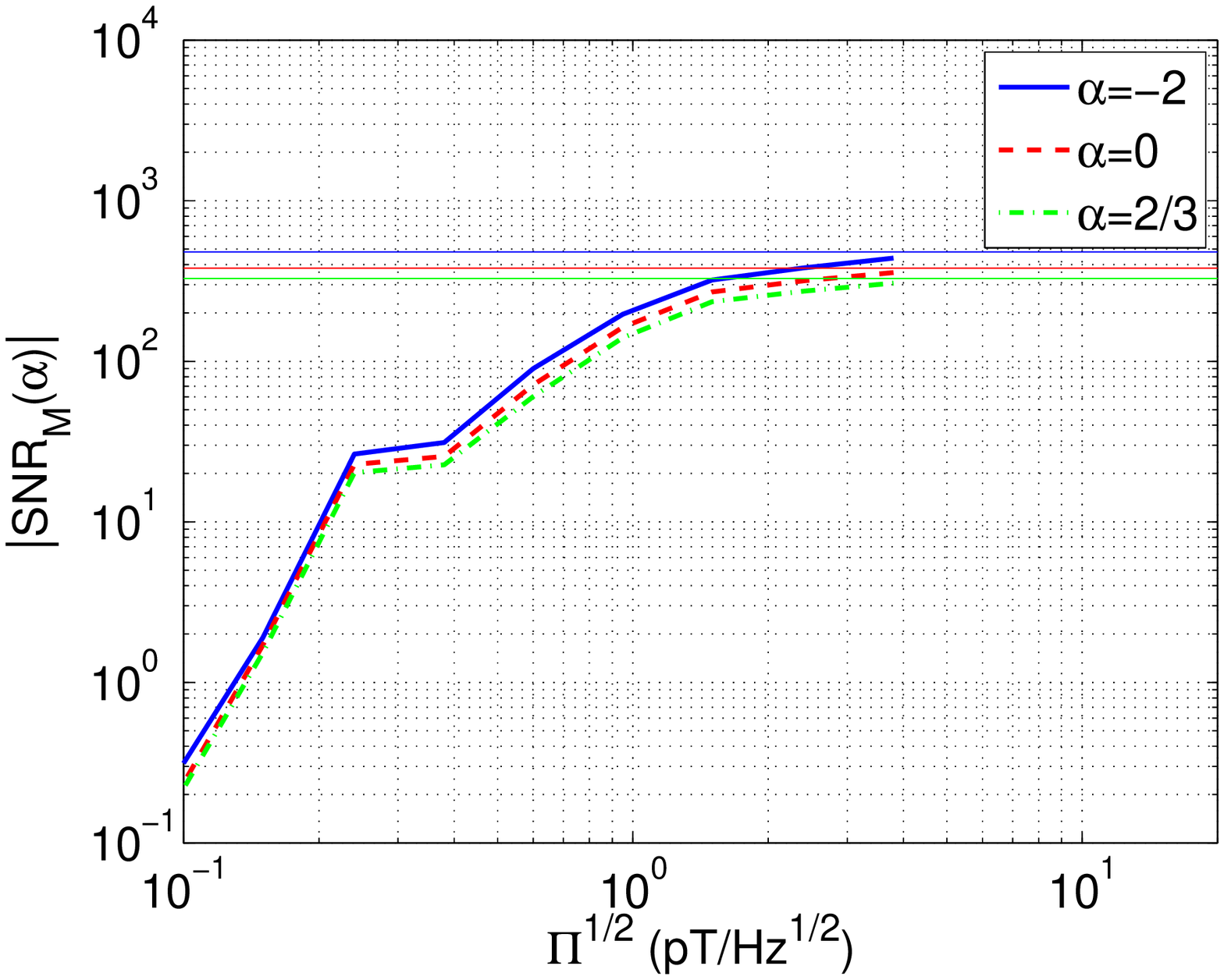, width=3.2in}
    % corr_mag_strain_noise.eps is made with aligo.m.  harvest.eps is made with
    % harvest.eps is made with new_harvest.eps (USING JUST ONE YEAR OF DATA)
    % example_rho.m is made with aligo.m.
    % snr_vs_alpha is made with snr_vs_alpha.m.
    %   ...which reads in archive/new_harvest files, created by new_harvest.m
    %   ...using run_aligo
  \end{tabular}
  \caption{
    Top-left: amplitude density spectra for uncorrelated (red) and correlated strain noise (cyan).
    The dashed red line shows the reduction in uncorrelated noise obtained through $\unit[1]{yr}$ of integration using a (typical~\cite{stoch-S5}) frequency bin width of $\delta f=\unit[0.25]{Hz}$.
    The blue shows a single realization of correlated noise.
    Top-right: coherence spectra showing the contaminated strain channels (blue), the somewhat cleaner spectrum obtained by Wiener-like filtering (black), and the spectrum obtained with true Wiener filtering (magenta).
    The Wiener-filtered spectrum is significantly cleaned, but measurable residual contamination remains.
% June 16:
%    For this example, $(\kappa,\beta)=(1,2.67)$ (see, Eq.~\ref{eq:kappa_beta}) and $\Pi^{1/2}=\unit[0.38]{pT\, Hz^{-1/2}}$.
    For this example, $(\kappa,\beta)=(2,2.67)$ (see, Eq.~\ref{eq:kappa_beta}) and $\Pi^{1/2}=\unit[0.15]{pT\, Hz^{-1/2}}$.
    Again, we assume $\unit[1]{yr}$ of integration time.
% June 16:
%    Bottom-left: witness signal-to-noise ratio assuming $\Pi^{1/2}=\unit[0.24]{pT\, Hz^{-1/2}}$.
    Bottom-left: witness signal-to-noise ratio assuming $\Pi^{1/2}=\unit[0.15]{pT\, Hz^{-1/2}}$.
    This noise floor is sufficient to remove detectable correlated noise.
    The $\unit[60]{Hz}$ electronic line has been notched.
    Bottom-right: the expected signal-to-noise ratio $|\text{SNR}_M(\alpha)|$ (Eq.~\ref{eq:SNRM}) from correlated noise as a function of the witness noise $\Pi^{1/2}$.
    The different colors show how the results vary for different gravitational-wave models.
    The dashed lines show dependence on $\Pi^{1/2}$ whereas the solid lines are the asymptotic values obtained when $\Pi^{1/2}\rightarrow\infty$ (no subtraction).
    All plots use the magnetic cross-power measurements from~\cite{schumann_ligo}.
  }
  \label{fig:aligo}
\end{figure*}

% June 16:
%In Fig.~\ref{fig:aligo}a, we show the amplitude spectral density for uncorrelated strain noise (red) and correlated strain noise from Schumann resonances (cyan) for parameters $(\kappa,\beta)=(1,2.67)$.
In Fig.~\ref{fig:aligo}a, we show the amplitude spectral density for uncorrelated strain noise (red) and correlated strain noise from Schumann resonances (cyan) for parameters $(\kappa,\beta)=(2,2.67)$.
The dashed red curve shows the uncorrelated noise (with $\delta f=\unit[0.25]{Hz}$ bins) achieved after one year of integration; see~\cite{locus}.
(Correlated noise is not reduced by integration.)
This choice of $(\kappa,\beta)$ corresponds to a realistic level of coupling.
The peak at $f=\unit[60]{Hz}$ is electronic in origin, and so we apply a frequency notch to exclude it from the analysis described below.

% June 16;
%In Fig.~\ref{fig:aligo}b, we show a coherence spectrum obtained for the same choice of $(\kappa,\beta)$ and assuming a witness noise of $\Pi^{1/2}=\unit[0.38]{pT\,Hz^{1/2}}$ and an integration time of $\unit[1]{yr}$.
In Fig.~\ref{fig:aligo}b, we show a coherence spectrum obtained for the same choice of $(\kappa,\beta)$ and assuming a witness noise of $\Pi^{1/2}=\unit[0.15]{pT\,Hz^{-1/2}}$ and an integration time of $\unit[1]{yr}$.
% June 16:
%Wiener filtering is able to produce a clean-looking coherence spectrum (purple), but there is still observable correlated noise present (detectable with a broadband statistic as in Eq.~\ref{eq:cohprime}; more on this below).
Wiener filtering is able to produce a clean-looking coherence spectrum (purple).

% June 16:
%In Fig.~\ref{fig:aligo}c, we plot $\rho_w(f)$ for the case of  $\Pi^{1/2}=\unit[0.24]{pT\,Hz^{1/2}}$, which, we show below, provides sufficient sensitivity to remove the correlated noise.
In Fig.~\ref{fig:aligo}c, we plot $\rho_w(f)$ for the case of  $\Pi^{1/2}=\unit[0.15]{pT\,Hz^{-1/2}}$, which, we show below, provides sufficient sensitivity to remove the correlated noise.
% June 16:
%Thus, for $(\kappa,\beta)=(1,2.67)$, we estimate that a witness of $\Pi^{1/2}=\unit[0.24]{pT\,Hz^{1/2}}$ [corresponding to $\rho_w(f)\approx1$--$4$] is required to subtract correlated noise in $\unit[1]{yr}$ of Advanced LIGO data at design sensitivity.
Thus, for $(\kappa,\beta)=(2,2.67)$, we estimate that a witness of $\Pi^{1/2}=\unit[0.15]{pT\,Hz^{-1/2}}$ [corresponding to $\rho_w(f)\approx2$--$6$] is required to subtract correlated noise in $\unit[1]{yr}$ of Advanced LIGO data at design sensitivity.

In the following subsection, we perform simulations for a range of witness noise $\Pi^{1/2}=0.1$--$\unit[4]{pT\,Hz^{-1/2}}$.
We record $\text{SNR}_M(\alpha)$ for three values of $\alpha$: $\alpha=0$ (expected for a cosmological source~\cite{stoch-S5}), $\alpha=2/3$ (expected for an astrophysical background of binary coalescences), and $\alpha=-2$ (chosen conservatively to emphasize the low frequency range responsible for the worst contamination).

\subsection{Results}
First, to characterize the extent of the challenge before us, we recall the results of Tab.~\ref{tab:contamination} showing $\text{SNR}_M(\alpha)$ predicted if Advanced LIGO does not employ any subtraction whatsoever.
% June 16:
%$\text{SNR}_M(\alpha)$ may range from $\gtrsim24$--$270$.
$\text{SNR}_M(\alpha)$ may range from $\gtrsim24$--$470$.

\begin{figure*}[hbtp!]
  \begin{tabular}{cc}
    \psfig{file=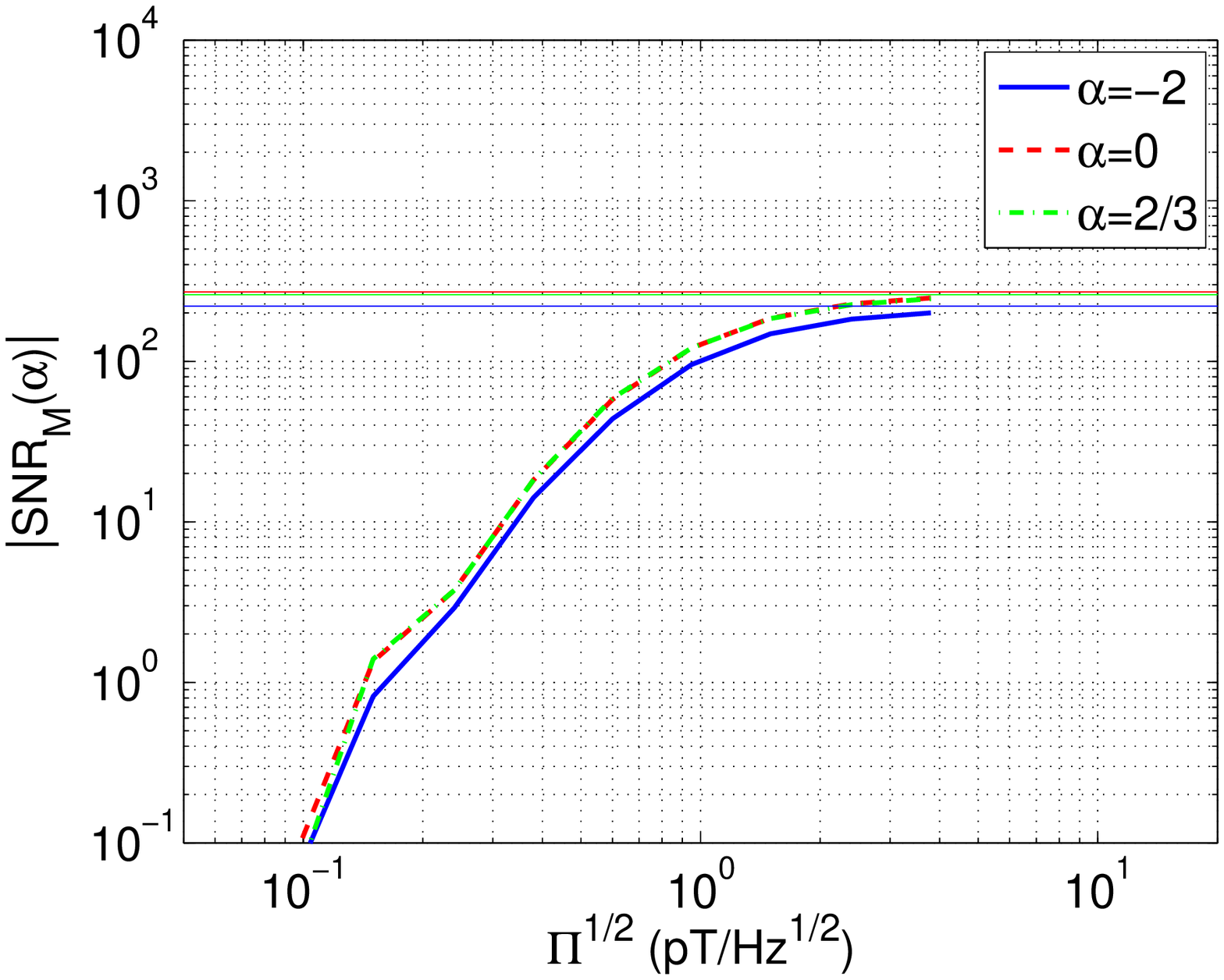, width=3.2in} &
    \psfig{file=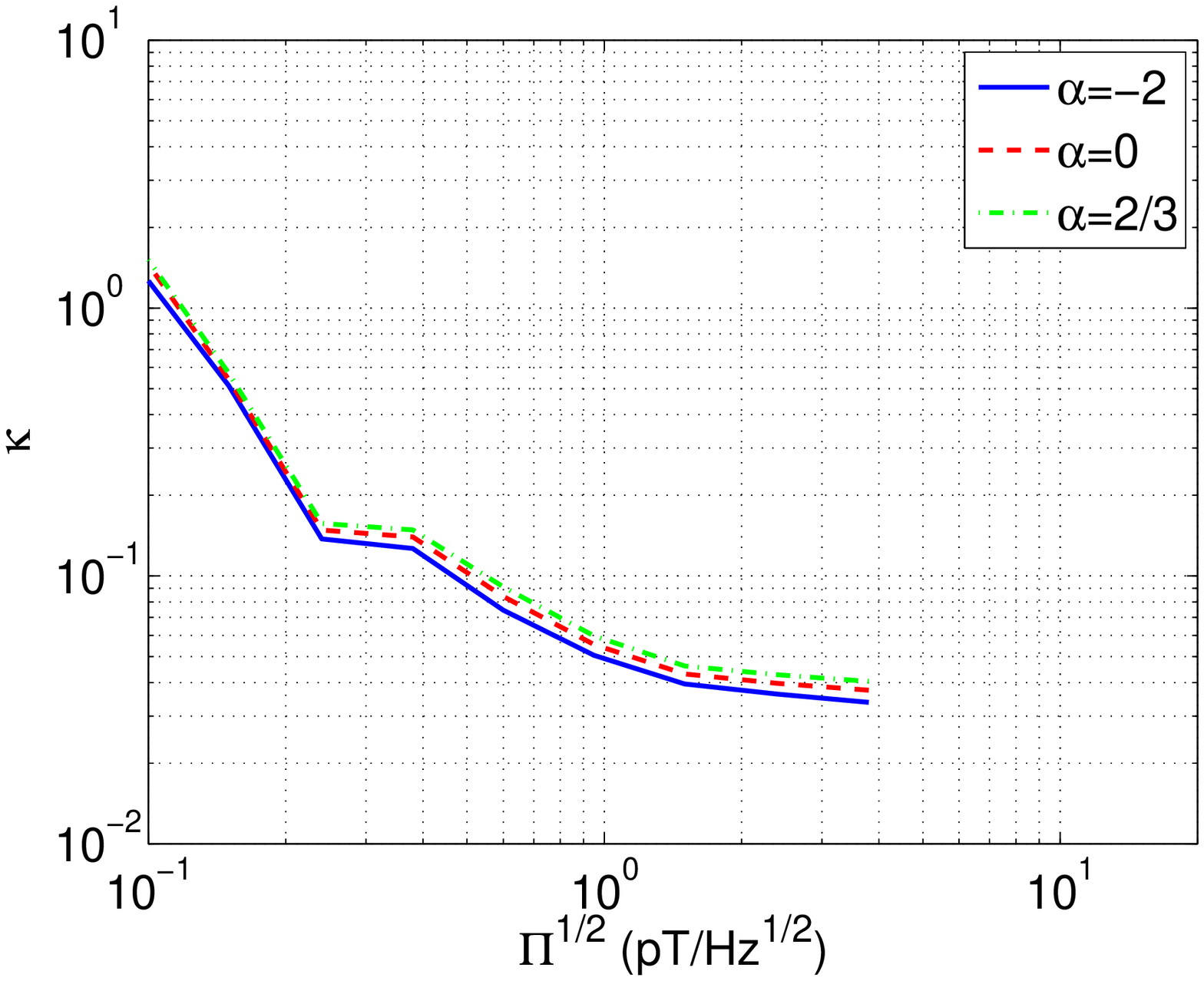, width=3.2in}
    % snr_vs_alpha_v2.eps is made with snr_vs_alpha_test.m.
    % kappa_vs_alpha.eps is made with kappa_vs_alpha.m.
  \end{tabular}
  \caption{
    Left: $|\text{SNR}_M(\alpha)|$ vs $\Pi^{1/2}$ (as in Fig.~\ref{fig:aligo}d) except with the $\unit[3]{mm}$ angular coupling scenario: $(\kappa,\beta)=(0.75,1.74)$.
% a steeper coupling function: $\kappa=6.4$ and $m=5.32$.
    The different colors indicate different models (different values of $\alpha$).
    Right: the required coupling $\kappa$ (given $\beta=2.67$) as a function of witness noise $\Pi^{1/2}$ required to reduce the correlated noise to a level $\text{SNR}_M(\alpha)\lesssim0.5$.
    The different colors indicate different models.
    Both plots use the magnetic cross-power measurements from~\cite{schumann_ligo}.
  }
  \label{fig:kappa}
\end{figure*}

In Fig.~\ref{fig:aligo}d, we show how $\text{SNR}_M(\alpha)$ decreases with decreasing witness noise $\Pi^{1/2}$.
% June 16:
%When $\Pi^{1/2}\lesssim\unit[0.1]{pT\, Hz^{-1/2}}$, the subtraction is sufficient to essentially eliminate correlated noise, no matter the spectral index.
When $\Pi^{1/2}\gtrsim\unit[2]{pT\, Hz^{-1/2}}$, the expected improvement from subtraction is negligible.
Thus, there is about one order of magnitude of witness noise $\Pi^{1/2}$ over which the efficacy of subtraction changes from effective to ineffective.
In Fig.~\ref{fig:kappa}a, we show the same $\text{SNR}_M(\alpha)$ vs $\Pi^{1/2}$ for the $x=\unit[3]{mm}$ angular coupling scenario: $(\kappa,\beta)=(0.75,1.74)$.
The overall shape is similar to Fig.~\ref{fig:aligo}d, which suggests that these results are somewhat robust to details about the shape of the coupling function.
The required witness noise is $\approx\unit[0.1]{pT\,Hz^{-1/2}}$.

It is interesting to consider how the results change when the coupling constant $\kappa$ is varied with a fixed spectral index $\beta$ and a fixed witness noise $\Pi^{1/2}$.
This is the type of change that one might expect from commissioning work to reduce magnetic coupling.
We find that
\begin{equation}
  \text{SNR}_M(\alpha|\Pi^{1/2}) \propto \kappa^2 ,
\end{equation}
i.e., changing $\kappa$ causes the curve in Fig.~\ref{fig:aligo}d to move up or down.
Thus, the expected contamination from correlated noise can be straightforwardly estimated by scaling with the appropriate factor of $\kappa^2$.

Following this logic, we can turn the question around and ask: what coupling is required in order to observe negligible contamination, e.g., $\text{SNR}_M(\alpha)<0.5$, given a witness noise of $\Pi^{1/2}$?
The answer to this question is shown in Fig.~\ref{fig:kappa}.
% June 16:
%We find that, given $\unit[1]{yr}$ of integration, and assuming $\alpha=2.67$, correlated noise can be reduced to manageable levels {\it without any subtraction whatsoever} if the coupling constant can be reduced to $\kappa\lesssim 7\times10^{-2}$.
We find that, given $\unit[1]{yr}$ of integration, and assuming $\alpha=2.67$, correlated noise can be reduced to manageable levels {\it without any subtraction whatsoever} if the coupling constant can be reduced to $\kappa\lesssim 3\times10^{-2}$.

\section{Contingency plans}\label{contingency}
In this section we consider the question of how to proceed in the event that subtraction is not sufficient to remove all traces of correlated noise.
First, though, we must address a related question: how do we judge if subtraction was successful?
We propose the following procedure.

First, construct a correlated noise budget as we have here.
This requires an estimate of the coupling function, which may be carried out using magnetic injection coils.
We expect the resultant coupling function to be uncertain to a factor of $\approx$$2$, which is woefully inadequate to use for subtraction (see the discussion of a priori filtering above), but is good enough for estimating the extent of correlated noise.
The noise budget also relies on coherence measurements of magnetic fields at the detector sites~\cite{schumann_ligo}.
Combining all the measurements, the correlated noise budget is given by
\begin{equation}
  H_M(f) \approx k |r_1(f)| \, |r_2(f)| \, 
  |\overline{\tilde{m}_1^*(f) \tilde{m}_2(f)} | .
\end{equation}
Using this noise budget, one can perform a numerical study as we have here to estimate $\text{SNR}_M(\alpha)$; see Fig.~\ref{fig:aligo}d.

If $\text{SNR}_M(\alpha)\ll1$, then we expect the residual correlated noise to be small enough to ignore.
Of course, it will be prudent to perform cross checks; an apparent gravitational-wave signal with the shape of the Schumann spectrum ought to arouse suspicion.
If, on the other hand, $\text{SNR}_M(\alpha)\gtrsim1$, then we must be prepared to inflate our error bars to account for the systematic effect of correlated noise.
Unfortunately, this may be necessary even if no significant correlated noise is present since the correlated noise budget must be constructed conservatively.

What is the best way to inflate the error bar to account for correlated noise?
For the sake of pedagogy, we will start with a naively simple solution and work toward the optimal solution, which minimizes the total measurement uncertainty: systematic + statistical.

The simplest naive solution is to add the broadband bias  with the broadband statistical uncertainty in quadrature to form a total uncertainty:
\begin{equation}
  \sigma^\text{tot}_\alpha = \sqrt{\left[\text{SNR}_M^2(\alpha) + 1\right]\sigma_\alpha^2} .
\end{equation}
% June 16:
%For the realistic example considered above ($\alpha=0$, $\kappa=1$, $\beta=2.67$), $\sigma_\alpha^\text{tot}=95\sigma_\alpha$, which represents a disastrous loss of sensitivity.
For the realistic example considered above ($\alpha=0$, $\kappa=2$, $\beta=2.67$), $\sigma_\alpha^\text{tot}=95\sigma_\alpha$, which represents a disastrous loss of sensitivity.

Fortunately, we can do better.
We can take advantage of our knowledge of the correlated noise spectral shape to minimize its impact on stochastic searches.
It is clear, e.g., from Fig.~\ref{aligo}b that the contamination is worst at low frequencies.
Thus, one can introduce a low-frequency cut-off $f_\text{min}$ to Eq.~\ref{eq:broadband} such that
\begin{equation}
  \begin{split}
    \widehat\Omega_M(\alpha) & = 
    \int_{f_\text{min}}^{f_\text{max}} df \, \widehat\Omega_M(f|\alpha) \, \sigma^{-2}(f|\alpha) \Big/ 
\int_{f_\text{min}}^{f_\text{max}} df \, \sigma^{-2}(f|\alpha) \\
    \sigma(\alpha) & = \left[ \int_{f_\text{min}}^{f_\text{max}} df \, \sigma^{-2}(f|\alpha) \right]^{-1/2} .
  \end{split}
\end{equation}
Then, $f_\text{min}$ can be tuned to minimize the total uncertainty.

\begin{figure*}[hbtp!]
  \begin{tabular}{cc}
    \psfig{file=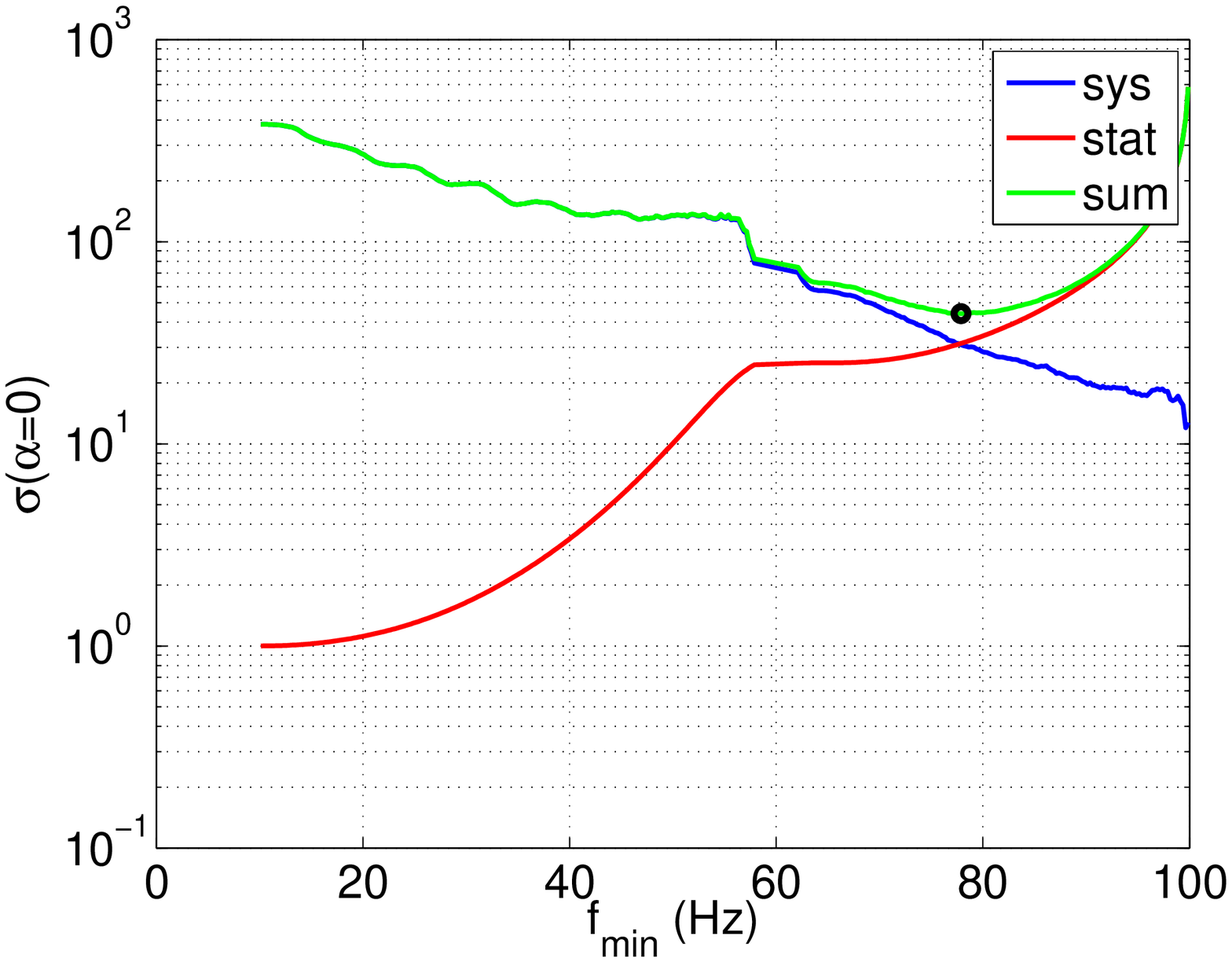, width=3.2in} & 
    \psfig{file=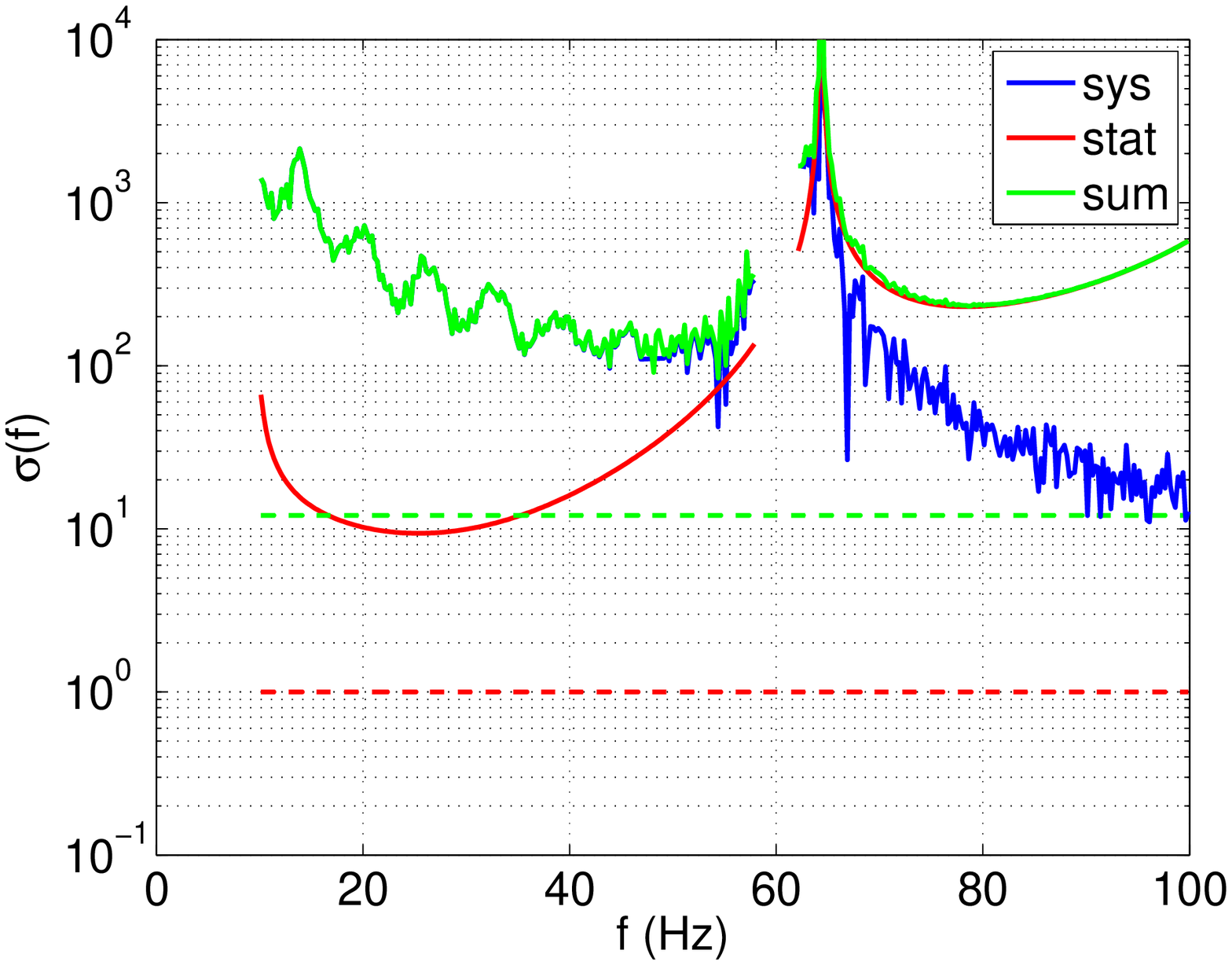, width=3.2in}
    % These plots are both made with noisebudget.m.
  \end{tabular}
  \caption{
    Left: systematic error (blue), statistical uncertainty (red), and their quadrature sum (green) as a function of the low frequency bound for the integrals in Eqs.~\ref{eq:broadband}.
% June 16:
%    The total error is minimized at $f_\text{min}=\unit[69]{Hz}$, which is marked with a black circle.
    The total error is minimized at $f_\text{min}=\unit[78]{Hz}$, which is marked with a black circle.
    Right: narrowband systematic error (blue), statistical uncertainty (red), and their quadrature sum (green).
    The solid lines show the uncertainty for many $\unit[0.25]{Hz}$-wide bins.
    The dashed lines show the broadband uncertainty obtained by the optimal combination of every frequency bin.
% June 16:
%    The broadband uncertainty with correlated noise is $4.2\times$ larger than the broadband uncertainty with no correlated noise.
    The broadband uncertainty with correlated noise is $12\times$ larger than the broadband uncertainty with no correlated noise.
% June 16:
%    In both plots, we assume $\alpha=0$, $\kappa=1$, and $\beta=2.67$.
    In both plots, we assume $\alpha=0$, $\kappa=2$, and $\beta=2.67$.
    Also, the normalization is such that $\sigma_0=1$.
    Finally, both plots use the magnetic cross-power measurements from~\cite{schumann_ligo}.
  }
  \label{fig:contingency}
\end{figure*}

Fig~\ref{fig:contingency}a shows the systematic error (bias) from correlated noise (blue), the statistical uncertainty (red), and their quadrature sum as a function of $f_\text{min}$.
% June 16:
%By selecting $f_\text{min}\approx\unit[69]{Hz}$, we can minimize the total uncertainty such that $\sigma^\text{tot}_\alpha\approx28\sigma_\alpha$.
By selecting $f_\text{min}\approx\unit[78]{Hz}$, we can minimize the total uncertainty such that $\sigma^\text{tot}_\alpha\approx44\sigma_\alpha$.

While a low-frequency cutoff is a better solution than naively adding the broadband bias to the statistical uncertainty in quadrature, it is still not the optimal solution, and we can do better yet.
The optimal solution is to combine the statistical and systematic uncertainty for each frequency bin individually {\it before} integrating over (the entire) frequency band.
This method ensures that we weight each frequency bin according to both the statistical and systematic uncertainty.

The total uncertainty associated with each frequency bin is:
\begin{equation}
  \sigma^\text{tot}(f|\alpha) = \sqrt{\widehat{\Omega}_M^2(f|\alpha) + \sigma^2(f|\alpha)} ,
\end{equation}
and the integrated uncertainty is given by
\begin{equation}
  \sigma^\text{tot}(\alpha) = \left( \int df \left[\sigma^\text{tot}(f|\alpha)\right]^{-2} \right)^{-1/2} .
\end{equation}

In Fig.~\ref{fig:contingency}b, we plot the systematic (blue), statistical (red), and total error (green) in each $\unit[0.25]{Hz}$ frequency bin as a function of frequency.
We also show as dashed lines the integrated broadband values for $\sigma_0^\text{tot}$ assuming purely statistical error (red) and including systematic error (green).
The normalization has been chosen so that $\sigma_0=1$.

% June 16
%Using ``optimal re-weighting,'' the loss of sensitivity is modest: $\sigma^\text{tot}_0=4.2\sigma_0$.
Using ``optimal re-weighting,'' the loss of sensitivity is modest: $\sigma^\text{tot}_0=12\sigma_0$.
This somewhat surprising result can be understood as follows.
While the correlated noise is very large over parts of the band, there are significant regions where it is small, and so the overall sensitivity is not strongly affected so long as regions of high contamination are appropriately re-weighted.

For the sake of completeness, we repeat the calculation using the other coupling from Fig.~\ref{fig:noisebudget}.
The results are given in Tab.~\ref{tab:reweight}.
The loss of sensitivity due to coupling through the angular degree of freedom strongly depends on the beam offset $x$.

\begin{table}
  \begin{tabular}{|c|c|}
    \hline
    {\bf coupling} & {\bf $\sigma^\text{tot}(\alpha=0)/\sigma(\alpha=0)$} \\\hline
    angular $x=\unit[1]{mm}$ & $2.5$ \\\hline
    angular $x=\unit[3]{mm}$ & $17$ \\\hline
% June 16: Robert's post-processing factor of two
%    length & $4.2$ \\\hline
    length & $12$ \\\hline
    % from noisebudget.m.
  \end{tabular}
  \caption{
    The expected loss of sensitivity for an $\alpha=0$ signal using $\unit[1]{yr}$ of LIGO Hanford-Livingston data at design sensitivity.
    The first column describes different coupling scenarios; see Fig.~\ref{fig:noisebudget}.
    The second gives the relative increase in the error bar.
  }
  \label{tab:reweight}
\end{table}

While optimal re-weighting can be used to minimize the impact of correlated noise on searches for {\it broadband} signals, it provides no help for cross-correlation searches for narrowband signals.
In particular, the machinery of the stochastic search---operating in a radiometer mode---has been applied with great success to search for gravitational waves from point sources such as the low-mass X-ray binary, Sco X-1~\cite{radiometer,sph_results}.
% June 16:
%Based on Fig.~\ref{fig:contingency}b, we estimate that, without successful subtraction, correlated noise at $f\lesssim\unit[45]{Hz}$ may increase the total uncertainty by a factor of $\gtrsim$$10$ in energy density, or equivalently, a factor of $\gtrsim$$3$ decrease in strain sensitivity (reducing the visible volume of the search by a factor of $\approx$$30$).
Based on Fig.~\ref{fig:contingency}b, we estimate that, without successful subtraction, correlated noise at $f\lesssim\unit[70]{Hz}$ may increase the total uncertainty by a factor of $\gtrsim$$60$ in energy density, or equivalently, a factor of $\gtrsim$$8$ decrease in strain sensitivity (reducing the visible volume of the search by a factor of $\approx$$460$).
The only means of preventing this deleterious outcome is through successful subtraction.

\section{Conclusions}\label{conclusions}
When searching for the stochastic background, advanced gravitational-wave detectors such as LIGO and Virgo must contend with correlated noise from Schumann resonances and possibly other geomagnetic phenomena.
Wiener filtering with magnetometers can, in principle, be used to reduce correlated noise in stochastic searches.
The success or failure of Wiener-filter subtraction depends crucially on the witness signal-to-noise ratio: the relative amplitude of Schumann resonance fields to noise, both from the magnetometer itself and from local magnetic fields.
It may be difficult to achieve the $\approx$$0.1$--$\unit[0.2]{pT\,Hz^{-1/2}}$ noise level thought necessary to completely remove correlated noise.
However, we have strived to provide guidance in this endeavor by highlighting some of the subtleties and pitfalls of correlated noise subtraction.

Given the challenge of effective subtraction, we stress that any reduction in magnetic coupling will pay immediate dividents in a reduction in correlated noise.
Similarly, it is crucially important to minimize the beam offset (from the axis of rotation) in order to reduce correlated noise coupling through angular degrees of freedom.

If residual correlated noise remains after subtraction, all is not lost.
We described how a conservative correlated noise budget can be used to re-weight the integration over frequency bins performed in a stochastic search, which, in turn, dilutes the correlated noise.
% June 16:
%For a realistic case of coupling, and assuming an isotropic search for a $\Omega_\text{gw}(f)=\text{const}$ source with $\unit[1]{yr}$ of data from LIGO-Hanford cross-correlated with LIGO-Livingston, we find that optimal re-weighting increases the uncertainty by a modest factor of $\approx$$4.4\times$.
For a realistic case of coupling, and assuming an isotropic search for a $\Omega_\text{gw}(f)=\text{const}$ source with $\unit[1]{yr}$ of data from LIGO-Hanford cross-correlated with LIGO-Livingston, we find that optimal re-weighting increases the uncertainty by a more modest factor of $\approx$$12\times$.
% June 16:
%However, cross-correlation searches for narrowband sources with $f\lesssim\unit[45]{Hz}$ will be significantly and adversely affected if correlated noise cannot be effectively subtracted or otherwise reduced.
However, cross-correlation searches for narrowband sources with $f\lesssim\unit[70]{Hz}$ will be significantly and adversely affected if correlated noise cannot be effectively subtracted or otherwise reduced.

There is much future work to be done.
Field studies in and around the LIGO-Virgo sites can be carried out to determine if there are suitable magnetically quiet locations for magnetometer stations.
Low-noise magnetometers can be evaluated in order to achieve the lowest possible noise floor.
Preliminary magnetometer data can be used to test the efficacy of different subtraction schemes.
Experimental tests will reveal difficult-to-predict effects, arising, e.g., from the non-stationarity of Schumann resonances themselves.

\begin{acknowledgments}
  We gratefully acknowledge the use of environmental monitoring data from the LIGO experiment.
  We thank Grant Meadors, Stan Whitcomb, David Yeaton-Massey, and Shivaraj Kandhasamy for helpful discussions.
  ET is a member of the LIGO Laboratory, which is supported by funding from United States National Science Foundation.
  LIGO was constructed by the California Institute of Technology and Massachusetts Institute of Technology with funding from the National Science Foundation and operates under cooperative agreement PHY-0757058.
  NC and RS are supported by NSF grants PHY-1204371 and PHY-0855686 respectively.
  AE is supported by NSF grants PHY-0905184 and PHY-1205882.
  This paper has been assigned LIGO document number P1400069.
\end{acknowledgments}

\bibliography{wsubtract}

\end{document}